\documentclass[11pt]{article}
\usepackage{latexsym,cite,amssymb}
\makeatletter

\@addtoreset{equation}{section}
\makeatother

\newcommand{\half}{{{\textstyle\frac{1}{2}}}}

\newcommand{\be}{\begin{equation} }
\newcommand{\ee}{\end{equation} }
\newcommand{\ba}{\begin{array} }
\newcommand{\ea}{\end{array} }

\newcommand\cE{{\cal E}}
\newcommand\cH{{\cal H}}

\newcommand\cJ{{\cal J}}
\newcommand\cL{{\cal L}}

\newcommand\cO{{\cal O}}
\newcommand\cQ{{\cal Q}}
\newcommand\cT{{\cal T}}
\newcommand\alphap{{\alpha^{\prime}}}

\newcommand{\so}{\mbox{so}}
\newcommand{\SO}{\mbox{SO}}
\newcommand\tr{{\rm tr}}
\newcommand\Tr{{\rm Tr}}
\newcommand\const{{\upsilon}}
\newcommand\rmd{{\rm d}}

\newcommand\para{g_{\hat{s}}}
\newcommand\phiX{{\small{\phi}}{(X)}}
\newcommand\Et{\cE_{t}}
\newcommand\Ep{\cE_{\phi}}
\newcommand\Epj{\cE_{\phi^{j}}}

\newcommand\outh{{\rm \scriptscriptstyle{out}}}
\newcommand\inh{{\rm\scriptscriptstyle{in}}}

\newcommand\THag{{T_{\scriptscriptstyle{\rm Hag.}}}}
\newcommand\Tbh{T_{\scriptscriptstyle{\rm BH}}}
\newcommand\Ncut{N_{\scriptscriptstyle{\rm cut}}}
\newcommand\rc{{\rm c}}
\newcommand\Einstein{{}}

\begin{document}
\begin{titlepage}
\title{\vskip -60pt
{\small
\begin{flushright}
hep-th/0603243\\
MPP-2006-37
\end{flushright}}
\vskip 20pt
Matrix models  for
D-particle dynamics and\\
the  string/black hole transition \\ ~~~\\~~~ }
\author{Johanna Erdmenger,   Jeong-Hyuck Park and Corneliu Sochichiu}
\date{}
\maketitle
\vspace{-1.0cm}
\begin{center}
~~~\\
\textit{Max-Planck-Institut f\"{u}r Physik,  F\"{o}hringer Ring 6,
80805  M\"{u}nchen,  Germany}\\
~~~\\
{\small Electronic correspondence: {{{jke, park, sochichi@}}mppmu.mpg.de}}\\
~~~\\
~~~\\
\end{center}
\begin{abstract}
\noindent  For a generic  two-dimensional 0A string background, we
map  the Dirac-Born-Infeld action to a matrix model.  This is
achieved using a canonical transformation.     The action  describes
D0-branes in this background,  while the matrix model  has   a
potential which encodes  all the information of the background
geometry.
We apply this  formalism to  specific   backgrounds: For Rindler space,
we obtain a matrix model with an upside-down quadratic potential,
while for    $AdS_{2}$ space,  the potential is     linear.
Furthermore we analyze  the   black hole geometry  with RR flux. In
particular, we show that at the Hagedorn temperature, the resulting
matrix model coincides with the one for the linear dilaton
background. We interpret this  result as a realization of  the
string/black hole transition.
\end{abstract}
~\\
\thispagestyle{empty}
\end{titlepage}
\newpage


\section{Introduction and summary}
As is well known,  the end points  of open strings are always attached to  D-branes, and  the
understanding  of  D-brane dynamics can provide an alternative description of
 open strings. In particular, in  noncritical two-dimensional  0A string theory,
the spectrum of possible stable D-branes is   restricted   to D-particles.    Therefore
the \textit{particle dynamics of the D-branes
should represent   the two-dimensional
open strings themselves.}  This essentially amounts to the holographic  principle
for  two-dimensional string theories or the `$AdS_{2}/{\rm{CFT}_{1}}$'
correspondence~\cite{Maldacena:1997re,Aharony:1999ti}.
Throughout this paper, we take this point of view as our guiding principle and
provide evidence for its validity. \\

In the present paper, we derive   matrix models from the one-dimensional
Dirac-Born-Infeld (DBI) action with  Chern-Simons (CS) term  which describes
D0-branes in  two-dimensional 0A string backgrounds.
The resulting matrix model is ``non-relativistic" in the sense that
the kinetic term is simply the velocity squared.
All the information of the background geometry is encoded in the  potential.
Our map from  the DBI-CS action  to  a matrix model  corresponds to
a canonical transformation in classical mechanics.
No approximation such as  the low velocity limit is taken.
Essentially the canonical transformation is possible  since  the system is integrable,
in the sense that  there is only one dynamical variable
matching the only conserved quantity, \textit{i.e.~}energy.\\

We apply our general formalism  to a number of
known two-dimensional backgrounds, including  black hole backgrounds with  flux.
Within string theory, black holes are formed by  condensation of D0-branes.
We work in the regime where the pair interaction between D-particles is weak, and
may therefore be replaced by the interaction with a self-consistent background,
which  in the present  case  is  the black hole geometry. Therefore, we start with a
probe analysis of the Born-Infeld action in the background, and obtain a corresponding
``non-relativistic" action for a single particle. Then we generalize it to a gauged matrix model to describe
a gas of D-particles. We show that in the gauged matrix model
the pair interaction between particles is negligible, and hence the resulting
matrix model is consistent with our probe analysis in the self-consistent background.
Moreover, the gauging of the matrix model is consistent with the fact that
D-particles are identical. Therefore we
 identify \textit{the gauged matrix model} as an
 \textit{effective description of the black hole itself.}
This will be our second guiding principle.
The above interpretation is in the spirit of   understanding  black holes  in terms of
falling matter outside the horizon, \textit{i.e.~}in the causally connected region.
In this region the string coupling is bounded from above by the
value at the horizon which is given by the inverse of the flux. Then,
by taking a large value of flux, we can make quantum corrections
negligible~\cite{Olsson:2005en}. In this case,
the classical Born-Infeld analysis for a probe brane can be trusted. \\

We also consider the linear dilaton background. Unlike the black hole backgrounds,
in this case the string coupling is not bounded and the
Born-Infeld analysis may not be trusted in the strong coupling region.
In spite of this, the classical  trajectory of the Born-Infeld action agrees with
the conformal field theory result where the trajectory is obtained
from the peak of the quantum wave
function~\cite{Douglas:2003up,Lapan:2005qz,Klebanov:2003km}. There is however an
issue of localization, as discussed  in \cite{Kutasov:2005rr}.\\

We observe  that the matrix model
for the black hole background at the Hagedorn temperature
is identical to the one
for the linear dilaton background.
This is a nontrivial result, as  the two backgrounds are geometrically  distinct and
the expressions of the matrix variable in terms of  the Liouville coordinate
are  different, as we will see in  (\ref{LLDXphi}) and (\ref{HBHXphi}).
We interpret  our result as   the `string/black hole transition' at the Hagedorn temperature,
which was initially proposed  by Susskind for  generic  black
holes~\cite{Susskind:1993aa,Susskind:1993ws,Susskind:2005js}, elaborated  further in
\cite{Horowitz:1996nw}, and recently discussed  for  two-dimensional black holes in
\cite{Giveon:2005mi,Nakayama:2005pk,Kutasov:2005rr,Giveon:2005jv,Parnachev:2005qr}. According to this
proposal,
there is a one-to-one correspondence between
black hole and  string states.
Here we see an explicit realization of this physical picture:
At the Hagedorn temperature, where the string coupling vanishes,
the black hole matrix model coincides with the one for the linear dilaton background.
The former is the matrix model for D-particles in the black hole background,
which we assume to describe the black hole itself, while
the latter describes fundamental strings.
At the level of the D0 trajectory,  the transition was already observed
in a similar way in \cite{Nakayama:2005pk}. Here we describe the transition in terms of
matrix models, taking particular care of the normalization of the potential. \\

The fact that the matrix models for two distinct  backgrounds coincide
in the vanishing  string coupling regime suggests that this coincidence
is a more generic phenomenon.
This leads us to the conjecture that
\textit{in the vanishing  string coupling limit, the D0-brane dynamics or string theory itself
 becomes  independent of background,  such that the resulting matrix model is
 universally   given by the matrix model with an upside-down quadratic potential,
 of which the coefficient is fixed in  string units.}
 In other words, in  the vanishing string coupling limit
    the  matrix model does not refer to any specific background geometry.
  In principle, this can be  checked explicitly by calculating physical quantities,
  such as  the spectrum or  scattering amplitudes in different backgrounds.\\

Boundary states of D0-branes  have been earlier
studied in a coset conformal field theory at level
$\kappa$~\cite{Maldacena:1997cg,Nakayama:2005pk,Kutasov:2005rr}
(see also \cite{Kazakov:2000pm,Kutasov:2004dj,Karczmarek:2004bw,
Suyama:2004vk,Davis:2004xb,Danielsson:2004xf,Davis:2004xi,
Thompson:2003fz,Lapan:2005qz,Olsson:2005en}). This background is generically different
from ours, except at a single intersecting point  $\kappa={1/2}$,
which in our case corresponds to the black hole at the Hagedorn temperature.\\

The organization and the summary of the present paper are as follows:\\
In Sec. \ref{secBH},  we
review the exact black hole solutions in two-dimensional  0A string theory, and
consider  the near-horizon as well as two different  weak coupling limits.
We also set up our notations to
express the solutions in terms of the black hole temperature.\\

In Sec. \ref{MBH}, we construct the canonical transformation which maps  DBI-CS actions
for D-particles in a generic two-dimensional string background to  matrix models.
In the latter, the kinetic term is of canonical form as in non-relativistic mechanics,
while the potential is nontrivial and contains all the information  of the background
geometry. The potential naturally decomposes into  DBI part and CS part.
In particular, we establish the relation between the matrix variable and
the Liouville coordinate for a generic background.
We also analyze
the isometry of the background, if present,  and discuss its implication for
the corresponding matrix model. We derive a necessary and sufficient condition
for the existence of an isometry, and solve the Killing equations completely.
We classify  possible isometry groups and show
that they are  generically three-dimensional such as $\SO(1,2)$, $\SO(3)$ or the
 Poincar\'{e} group.
The isometry is inherited by the matrix model for the  DBI sector
and always  gives rise  to a  $\SO(1,2)$ symmetry. An important result is that
\textit{the isometry of the background geometry implies that
the DBI potential is  a polynomial, at most quadratic in the matrix variable.} This is
also consistent with  the earlier work \cite{Park:2005pz}, where it was shown that
only  such matrix models  allow for $\mbox{OSp}(1|2,{\bf{R}})$
supersymmetric extension.
We also show that the condition
for the isometry is equivalent to
the static equation of motion in Liouville field theory. This supports our proposal
that our matrix models provide an appropriate description of noncritical
two-dimensional 0A string theory.\\

Section  \ref{SimpleSec} contains  some simple applications.
We apply the general formalism to   Rindler space,
$AdS_{2}$  and the linear dilaton backgrounds,
and derive the corresponding matrix models.
For  Rindler space as well as the linear dilaton background,
the resulting matrix models have an upside-down quadratic potential.
For $AdS_{2}$  the potential  is linear.
The  results for Rindler and $AdS_{2}$ spaces  are
interpreted as the first order approximation of the description of
noncritical strings near the  non-extremal and extremal black hole horizons.\\

In section \ref{MBHsec} we  carry  out a detailed analysis of the
matrix models for the exact black hole geometry. Firstly, we show that the matrix variable is
positive semi-definite $X\geq 0$. Spatial infinity and  outer horizon correspond
to $X=0$ and $X=\infty$, respectively.
For the black hole at the Hagedorn temperature, we find that
the corresponding  matrix model coincides with the one we obtained
for the linear dilaton
background. This corresponds to a realization of the string/black hole transition.
Further, we conjecture that
in the vanishing string coupling limit, irrespective of the background geometry,
the matrix model is of  universal form.   We also consider
the black hole  near the  Hagedorn temperature, and derive
the corresponding matrix model with a nontrivial potential, involving an `arctangent' function.\\

We continue by  studying  the near-horizon geometries of
the extremal and non-extremal black holes by taking into account
corrections to the matrix model potential up to the string
length scale.    The resulting matrix model depends on the black hole temperature explicitly.
For the non-extremal black hole, it essentially agrees with the `deformed matrix model'
proposed by Jevicki and Yoneya~\cite{Jevicki:1993zg}.
On the other hand for the extremal black hole, the potential contains
 linear,  square root and logarithmic terms.\\

Beyond the 0A gravity background, we also consider the
black hole background in a coset conformal field theory at level $\kappa$.
When $\kappa={1/2}$, this geometry coincides with the one for the black hole
at the Hagedorn temperature.
We obtain the corresponding matrix model with   $\kappa$-dependent potential.
We identify the black hole temperature as  function of $\kappa$, and
find agreement with the conformal field theory result.\\

In the  appendix we review
how the $\SO(1,2)$ symmetry inherited from the isometry of the background geometry is
realized in the matrix model, following \cite{Park:2005pz}.
The $\SO(1,2)$ exists if the potential consists of
quadratic, linear and inverse square terms.
Especially when  the quadratic term is absent, the matrix model possesses $\SO(1,2)$
conformal symmetry.
This applies to  most of the  matrix models we obtain, in particular to the
Rindler, $AdS_{2}$, linear dilaton,   non-extremal black hole and CFT background
matrix models. However, for the extremal black hole  case, there are logarithmic
and square root contributions to the potential. \\

\section{Low energy effective action and  black hole geometry\label{secBH}}
The   low energy effective action - in the Wilsonian sense - for the
two-dimensional type 0A string theory  reads
\be
\displaystyle{S_{{\rm{0A}}}=S_{\rm{bulk}}+S_{\rm{brane}}+S_{\rm{int.}}\,.}
\ee
Here $S_{{\rm bulk}}$ is an action of two-dimensional gravity coupled to
a closed string tachyon $T$ and a pair of RR field strengths $F^{(+)}$ and  $F^{(-)}$.
Explicitly \cite{Douglas:2003up,Thompson:2003fz,Strominger:2003tm,Gukov:2003yp},
\be
\ba{ll}
S_{\rm{bulk}}=&\displaystyle{-\int{\rm d}^{2}x\sqrt{-g}\left[\frac{1}{2\kappa^{2}}e^{-2\Phi}\left(
\frac{8}{\alpha^{\prime}}+R+4\left(\nabla\Phi\right)^{2}-f_{1}(T)\left(\nabla T\right)^{2}+
f_{2}(T)+\cdots\right)\right.}\\
{}&{}\\
{}&\displaystyle{~~~~~~~~~~~~~~~~~~~~~~~~
\left.-\frac{2\pi\alphap}{4}\left(f_{3}(T)(F^{(+)})^{2}+
f_{3}(-T)(F^{(-)})^{2}\right)+\cdots\right]\,.}
\ea
\label{Sbulk}
\ee
As there are two different RR gauge potentials, there are also two distinct D0-branes,
${\rm{D}}0^{(+)}$ and ${\rm{D}}0^{(-)}$ \cite{Douglas:2003up}. Accordingly the action for the
D-branes,
$S_{\rm{brane}}$, decomposes into $S_{\rm{brane}}^{(+)}+S_{\rm{brane}}^{(-)}$,
each of which corresponds to a non-Abelian Dirac-Born-Infeld-like
action for relativistic particles,
\be
S_{\rm{brane}}^{(\pm)}=
\displaystyle{-\frac{1}{~\kappa\sqrt{\alphap}~}\int{\rm d}s\,``\Tr"\left(e^{-\Phi}\sqrt{-g_{\mu\nu}
\dot{x}^{(\pm)\mu}\dot{x}^{(\pm)\nu}}\right)\,.}
\label{Sbrane}
\ee
Here we write    ``$\Tr$"  with a double quotation mark, as
the precise form of the $2D$ 0A non-Abelian Dirac-Born-Infeld action is not known. However,
in the weak  coupling region  which we consider in the subsequent,
only the Abelian part contributes.

Finally, $S_{\rm{int.}}$ describes the interactions
between the bulk modes and the brane modes, and includes
 Chern-Simons-like  terms for the coupling of the
RR gauge potential to the D-particles,
\be
S_{\rm{int.}}=\sqrt{2}
\displaystyle{\int{\rm d}s\,\Tr\left(C^{(+)}_{\mu}\dot{x}^{(+)\mu}\right)
+\Tr\left(C^{(-)}_{\mu}\dot{x}^{(-)\mu}\right)\,+\,\cdots\,.}
\label{Sinteraction}
\ee
In particular, in the presence of  $N^{(+)}$ number of ${\rm{D}}0^{(+)}$ static branes,
the equation of motion for $C_{t}^{(+)}$ reads with $x^{0}\equiv t$, $x^{1}\equiv\phi$,
\be
\displaystyle{\pi\alphap\partial_{\phi}\left(f_{3}(T)F^{(+)}\right)=\sqrt{2}\sum_{n}^{N^{(+)}}
\delta(\phi-\phi_{n}(t))\,.}
\ee
Integrating this  over $\phi$, we see that the flux is quantized.
One may also ask  about  contributions  of the brane action (\ref{Sbrane})
to other equations of motion for the bulk NS fields. However, since we  focus on the weak
string coupling region  we can neglect these.\\

In \cite{Banks:1992xs,Berkovits:2001tg}
(see also \cite{Davis:2004xb,Danielsson:2004xf,Thompson:2003fz}),
 black hole solutions are obtained for the equations of motion derived from
 the bulk action (\ref{Sbulk})
for vanishing tachyon. Requiring the tachyon to vanish $(T\equiv 0)$ implies that
the  fluxes must be equal,\footnote{
The normalization of $S_{{\rm int.}}$ has been merely chosen to match the result of
\cite{Douglas:2003up}  where the flux is precisely the number of D-particles.}
\cite{Gukov:2003yp,Davis:2004xi},
\be
\ba{ll}
\displaystyle{(F^{(+)})^{2}=(F^{(-)})^{2}=\frac{Q^{2}}{~2\pi^{2}\alphap^{2}}\,,}
~~~~&~~~~Q=N^{(+)}=N^{(-)}\,.
\ea
\ee
The flux then plays the role of a negative cosmological constant, and
the solutions of the equations of motion of the bulk action (\ref{Sbulk})
are given by, after setting $2\kappa^{2}\equiv1$,
\be
\ba{lll}
\displaystyle{{\rm d}s^{2}=-l(\phi){\rm d}t^{2}+\frac{1}{\,l(\phi)}{\rm d}\phi^{2}\,,}~~&~~~
\displaystyle{
l(\phi)=1+e^{2\sqrt{2}\phi m_{s}}\!\left(\frac{\left(g_{s}Q\right)^{2}}{4\sqrt{2}\,\pi}
\phi m_{s}-\xi\right)\,,}\\
{}&{}\\
\displaystyle{\Phi(\phi)=\sqrt{2}\phi m_{s}+\Phi_{0}\,,}~~&~~~
\displaystyle{C^{(\pm)}_{t}(\phi)=-\frac{Q}{\sqrt{2}\pi\alphap}\,\phi+
C^{(\pm)}_{t}(0)\,,~~~~~~~~~
C^{(\pm)}_{\phi}=0\,.}
\ea
\label{BHsolution}
\ee
Here $m_{s}={\alphap}^{-\frac{1}{2}}$ is the string mass scale, $\xi$ is a
dimensionless  free parameter, and $\Phi_{0}$ is  the vev of the
dilaton which gives the string coupling at $\phi=0$,
\be
\displaystyle{g_{s}=e^{\Phi_{0}}\,.}
\ee
The horizon, $\phi_{h}$, is, by definition,  located at   $l(\phi_{h})=0$,
so that we can express the free parameter $\xi$  by the horizon,
\be
\displaystyle{\xi=e^{-2\sqrt{2}\phi_{h} m_{s}}
+\frac{\left(g_{s}Q\right)^{2}}{4\sqrt{2}\pi}\phi_{h}m_{s}}\,.
\ee
As a function of $\phi_{h}$, $\xi$ is bounded below,\footnote{The case
$\xi<\left.\xi\right|_{\rm{extremal}}$ corresponds to a  naked singularity such that
 classical gravity breaks down. We exclude this possibility.}
\be
\displaystyle{\xi\geq
\left.\xi\right|_{\rm{extremal}}=\frac{\left(g_{s}Q\right)^{2}}{\,16\pi\,}
\left[1-\ln\left(\frac{\left(g_{s}Q\right)^{2}}{16\pi}\right)\right]\,,}
\label{cextrem}
\ee
and generically there are two horizons, the inner horizon $\phi^{{\inh}}_{h}$,
and the outer horizon $\phi_{h}^{{\outh}}$,
satisfying $\phi_{h}^{{\outh}}\leq\phi_{h}^{{\inh}}$,
as the region $\phi\rightarrow-\infty$ corresponds to the
asymptotically flat spatial infinity.
When the bound on $\xi$ (\ref{cextrem}) is saturated, the horizon assumes
the critical value
\be
\displaystyle{\left.\phi_{h}\right|_{\rm{extremal}}=-\sqrt{\frac{\alphap}{8}}\ln
\left(\frac{\left(g_{s}Q\right)^{2}}{16{\pi}}\right)\,.}
\label{extremalh}
\ee
This corresponds to the extremal black hole, where the two horizons coincide.
In general, we have
\be
\phi_{h}^{{\outh}}~\leq\,\left.\phi_{h}\right|_{\rm{extremal}}\,\leq~
\phi_{h}^{{\inh}}\,.
\label{hhh}
\ee
Further, it is convenient to introduce the following parameters:
the distance to the horizon and the derivative of $l(\phi)$ at the horizon,
\be
\ba{ll}
u\equiv \phi-\phi_{h}\,,~~~~&~~~~
\displaystyle{\varepsilon:=\sqrt{\alphap}
\left.\partial_{\phi}l(\phi)\right|_{\phi=\phi_{h}}=2\sqrt{2}\left(
\frac{\left(g_{s}Q\right)^{2}}{16\pi}e^{2\sqrt{2}\phi_{h}m_{s}}-1\right)}\,.
\ea
\label{varepdef}
\ee
Note that the two horizons, the inner and the outer,
can be distinguished by the sign of $\varepsilon$, \textit{i.e.~}$\varepsilon^{\inh}>0$
for the inner horizon and $\varepsilon^{\outh} <0$ for the outer horizon.
The temperature of the black hole is then given by $\left|\varepsilon^{\outh}\right|$
at the outer horizon,
\be
\displaystyle{{\Tbh}:=
-\frac{1}{4\pi}\left.\partial_{\phi}l(\phi)\right|_{\phi=\phi_{h}^{{\outh}}}
=\frac{m_{s}}{4\pi}\,\left|\varepsilon^{{\outh}}\right|=
\frac{m_{s}}{\sqrt{2}\pi}\left(1-
\frac{\left(g_{s}Q\right)^{2}}{16\pi}e^{2\sqrt{2}\phi_{h}^{{\outh}}m_{s}}\right)}\,,
\label{temp}
\ee
which is bounded  by the Hagedorn temperature from above,
\be
\displaystyle{\Tbh\leq\THag=\frac{m_{s}}{\sqrt{2}\pi}\,.}
\label{Hagedorn}
\ee
Note that in all the expressions above,  the dependence on
the flux and the string coupling is through their product $g_{s}Q$ only.\\

Instead of $\Phi_{0}$ and $\xi$, we may take  $\Phi(\phi_{h})$ and  $\varepsilon$
as  alternative free parameters. Then with $\Phi(\phi)=\sqrt{2}u m_{s}+\Phi(\phi_{h})$,
we can rewrite the black hole solution (\ref{BHsolution}) as
\be
\ba{ll}
l(\phi)&\displaystyle{=1+\left[\left(2\sqrt{2}+\varepsilon\right)um_{s}-1\right]
e^{2\sqrt{2}um_{s}}}\\
{}&{}\\
{}&\displaystyle{=\varepsilon um_{s}+\left(4+2\sqrt{2}\varepsilon\right)
(um_{s})^{2}+\left(\frac{16\sqrt{2}}{3}+4\varepsilon\right)(um_{s})^{3}+\cO\left[(um_{s})^{4}
\right]\,.}
\ea
\label{lexact}
\ee
Note that the above formula holds for both the inner and outer horizons, \textit{i.e.~}the
same function
$l(\phi)$ allows for two different expressions. However, henceforth
we will focus on the region  outside the outer horizon
so that $\varepsilon^{\outh}\leq 0$ and $u\leq 0$. This ensures that, from (\ref{hhh}),
the string coupling outside the outer horizon  is bounded from above by the value
for the extremal black hole~\cite{Olsson:2005en}
\be
\displaystyle{
e^{\Phi\left(\phi_{h}^{\outh}\right)}\leq
e^{\Phi\left(\left.\phi_{h}\right|_{\rm{extremal}}\right)}=\frac{4\sqrt{\pi}}{Q}\,.}
\label{gbound}
\ee
Thus, for  large $Q$ the string coupling is small outside the outer horizon.\\

We now consider various  weak coupling limits
 as well as the near-horizon limits of the exact black hole solution
(\ref{BHsolution}).
\begin{enumerate}
\item {\bf{Rindler space:}}\\
For the non-extremal black hole solutions  $\Tbh\neq 0$,
the near  outer horizon geometry $u<0$ is  two-dimensional  Rindler
space from (\ref{lexact}),~\cite{Davis:2004xb}
\be
\displaystyle{{\rm d}s^{2}=4\pi\Tbh u\,{\rm d}t^{2}
-\frac{1}{\,4\pi\Tbh u}\,{\rm d}u^{2}\,.}
\label{Rindler}
\ee

\item  ${\mathbf{AdS_{2}:}}$ \\
For the extremal black hole, $\Tbh=0$,
 the   near-horizon geometry of the extremal black hole  is $AdS_{2}$,
\be
\displaystyle{{\rm d}s^{2}=-4(um_{s})^{2}{\rm d}t^{2}+
\frac{1}{\,4(um_{s})^{2}}{\rm d}u^{2}\,.}
\label{AdSBH}
\ee

\item {\bf{Black hole at the Hagedorn temperature:}}\\
Taking $g_{s}\rightarrow 0$   in (\ref{BHsolution})  with the choice
$\xi\neq 0$, the solution  (\ref{BHsolution})
reduces  to the metric for the black hole at the Hagedorn
temperature~\cite{Solodukhin:1995te},
\be
\ba{ll}
 \displaystyle{{\rm d}s^{2}=-\left(1-e^{2\sqrt{2}um_{s}}\right)
{\rm d}t^{2}+\frac{1}{1-e^{2\sqrt{2}um_{s}}}\,{\rm d}u^{2}\,,}~~~~&~~~~
\displaystyle{\Phi-\Phi(\phi_{h})=\sqrt{2}u m_{s}\,.}
\ea
\label{HagBHsol}
\ee

\item {\bf{Linear dilaton background:}}\\
Taking $g_{s}\rightarrow 0$   in (\ref{BHsolution}), this time
choosing  $\xi=0$
(or alternatively  in the asymptotically flat region $\phi\rightarrow -\infty$),
the solution  (\ref{BHsolution})  becomes the linear dilaton background,
\be
\ba{ll}
\displaystyle{{\rm d}s^{2}=-{\rm d}t^{2}+{\rm d}\phi^{2}\,,}~~~&~~~
\displaystyle{\Phi=\sqrt{2}\,\phi m_{s}+\Phi_{0}\,.}
\ea
\label{lineardilaton}
\ee
\end{enumerate}

Note that the string coupling given by
$\displaystyle{e^{\Phi}=g_{s}e^{\sqrt{2}\phi m_{s}}}$
grows as $\phi$ becomes large. One should bear
in mind that in the strong coupling region the gravity action (\ref{Sbulk}) is
subject to quantum corrections and the solution~(\ref{BHsolution}) may not be valid.
This restricts the validity of the linear dilation background~(\ref{lineardilaton}) to
$\phi\ll\sqrt{\alphap}|\ln g_{s}|$. However, in contrast to the linear dilaton
background, for the generic black hole solution we consider,
the string coupling is bounded from above (\ref{gbound}) outside of the outer horizon.
The strong coupling region is  inaccessible for  a fiducial observer.\\
~\\
~\\

In addition to  the above black hole solutions of the gravity action (\ref{Sbulk}),
we also consider `black hole' background
in a coset conformal field theory at level $\kappa$, studied in
\cite{Maldacena:1997cg,Nakayama:2005pk,Kutasov:2005rr}
\be
\ba{ll}
\displaystyle{\rmd
s^{2}=\kappa\alphap\left(-\tanh^{2}\!\rho\,\rmd\tau^{2}+\rmd\rho^{2}\right)\,,}
~~~~&~~~~\displaystyle{\Phi=-\ln\cosh\rho +\Phi_{h}\,.}
\ea
\label{kmetric}
\ee
When $\kappa={1/2}$, this coincides with the black hole background at
the Hagedorn temperature (\ref{HagBHsol}), while for generic $\kappa\neq{1/2}$,
it is not a classical solution  of the gravity action (\ref{Sbulk}).
The dimensionless coordinates $\tau$, $\rho$ are related to
our time and Liouville coordinates $t,u$ through, with $\alphap=m_{s}^{-2}$,
\be
\ba{ll}
tm_{s}=\sqrt{\kappa}\,\tau\,,~~~~&~~~~\sqrt{2}um_{s}=-\ln\cosh\rho\,,
\ea
\ee
such that the horizon lies at $\rho=0$ ($u=0$), and at the spatial infinity
$\rho=+\infty$ ($u=-\infty)$ we have the normalization $g_{tt}=-1$.
In terms of $t,u$,  (\ref{kmetric})
can be rewritten as
\be
\ba{ll}
 \displaystyle{{\rm d}s^{2}=-\left(1-e^{2\sqrt{2}um_{s}}\right)
{\rm d}t^{2}+\frac{2\kappa}{\,1-e^{2\sqrt{2}um_{s}}}\,{\rm d}u^{2}\,,}~~~~&~~~~
\displaystyle{\Phi=\sqrt{2}u m_{s}+\Phi(\phi_{h})\,.}
\ea
\label{kmetric2}
\ee
Hence when $\kappa={1/2}$, this reduces to (\ref{HagBHsol}). Further,
in \cite{Kutasov:2005rr} it was shown that
 the wave function in the conformal field theory does not spread over a region
 but has a sharp localization when  $\kappa={1/2}$.~\footnote{
 $\kappa$  is related to ``$Q$" in  \cite{Kutasov:2005rr}  by
$``Q"={\sqrt{2/\kappa}}$.}
This also justifies our classical analysis to take D-particle as a point particle.
Applying our matrix model analysis  to
this additional background will allow us to compare
with the conformal field theory  result.

~\newline
\section{Matrix model  for the two-dimensional DBI-CS action\label{MBH}}
\subsection{Non-relativistic formulation of the DBI-CS action}
We consider $N$ D-particles on  generic two-dimensional backgrounds, in view of considering
 the black hole solution (\ref{BHsolution}).
We take the string coupling $g_{s}$ to be small such that
the backreaction is negligible, and
the interaction among  the $N$ D-particles is switched off.
For weak  string coupling it is sufficient to
consider a single D-particle treated as a probe.  Its dynamics is
governed by  the Abelian Dirac-Born-Infeld action (\ref{Sbrane}) with
  Chern-Simons term  (\ref{Sinteraction}).\\

In a gauge where the  metric is diagonal  $g_{t\phi}\equiv 0$,
and the world line parameter is identified with the target space time, $s\equiv t$,
the  action  for D0-brane dynamics is of the general form
\be
S_{D0}=\displaystyle{\int{\rm d}t\left[\,\mp m_{s}\sqrt{\Et(t,\phi)^{2}\mp
\Ep(t,\phi)^{2}\dot{\phi}^{2}}\,+\,C_{t}(t,\phi)+C_{\phi}(t,\phi)\dot{\phi}\,\right]\,,}
\label{LDBIab}
\ee
where we take the `zweibein' for the effective metric
$G_{\mu\nu}^{\Einstein}\equiv e^{-2\Phi}g_{\mu\nu}$ to be
\be
\ba{ll}
\Et^{2}:=e^{-2\Phi}\left|g_{tt}\right|\,>\,0\,,~~~~&~~~~
\Ep^{2}:=e^{-2\Phi}\left|g_{\phi\phi}\right|\,>\,0\,.
\ea
\label{abdef}
\ee
The upper sign is for
the Minkowskian spacetime,  while the lower one is for the Euclidean space.
Henceforth we focus on the static backgrounds, \textit{i.e.~}no explicit time dependence in
the backgrounds,
$\Et(\phi)$, $\Ep(\phi)$, $C_{t}(\phi)$, $C_{\phi}(\phi)$
as in the black hole solution (\ref{BHsolution}). Consequently
 the last term in the action drops out, and the Hamiltonian reads
\be
\displaystyle{H_{D0}=H_{{\rm{DBI}}}-C_{t}(\phi)\,,}
\label{HD0}
\ee
where $H_{{\rm DBI}}$ is the Hamiltonian for the DBI action.
With the canonical momentum for $\phi$,
\be
\displaystyle{p_{\phi}
=\frac{m_{s}\dot{\phi}\,\Ep^{2}}{\sqrt{\Et^{2}\mp\Ep^{2}\dot{\phi}^{2}}}\,,}
\ee
it  reads
\be
\displaystyle{H_{{\rm DBI}}=\pm \,m_{s}\!\left(\left(\frac{1}{\Et}\right)^{2}
\mp\left(\frac{\dot{\phi}\Ep}{\Et^{2}}\right)^{2}\right)^{-\frac{1}{2}}\!
=\pm\,m_{s}\sqrt{\Et^{2}\pm \left(\frac{p_{\phi}\Et}{m_{s}\Ep}\right)^{\!2}}\,.}
\ee
~\\

We define a new variable $X$  by
\be
\displaystyle{X(\phi):=\frac{1}{\,\para}
\int_{\phi_{0}}^{\phi}{\rm d}\phi^{\prime}\,\frac{\Ep(\phi^{\prime})}{~
\Et(\phi^{\prime})^{2}}~,}
\label{Xdef}
\ee
satisfying $\dot{X}=\dot{\phi}\Ep/\Et^{2}$.
This will turn out to be the matrix model variable, and
Eq.(\ref{Xdef}) establishes the relation between
the matrix variable  and the Liouville coordinate for generic backgrounds, as we see below.
Here we introduce an dimensionless parameter $\para$ which is a priori
an arbitrary constant. However, the natural choice is to identify it with
the string coupling, such that  $X$ in (\ref{Xdef}) is independent of
the string coupling. Hence, near the horizon region of black holes we  set
\be
\para\equiv e^{\Phi(\phi^{\outh}_{h})}\,.
\label{ghats}
\ee
Since $\Et$, $\Ep$ are non-negative definite,
$X(\phi)$ is monotonically increasing with $\phi$.
In terms of the new coordinates, $(t,X)$, the effective metric is
\be
\displaystyle{e^{-2\Phi}{\rm d}s^{2}=e^{-2\Phi}g_{\mu\nu}{\rm d}x^{\mu}{\rm d}x^{\nu}=
\mp\Et^{2}{\rm d}t^{2}+\para^{2}\Et^{4}{\rm d}X^{2}\,.}
\ee

Now we   rewrite (\ref{HD0}) as
\be
\displaystyle{
0=\dot{X}^{2}\mp\frac{1}{~\para^{2}\Et(\phiX)^{2}}\pm\frac{m_{s}^{2}}{~
\para^{2}\left[\,H_{D0}+C_{t}(\phiX)\,\right]^{2}}\,.}
\label{0X}
\ee
From (\ref{0X}), it follows that the  D0 dynamics on the generic static backgrounds has an alternative dual
description as a non-relativistic  particle motion,
\be
\displaystyle{\cL_{{\rm nonrel.}}\,=\,\half\dot{X}^{2}\,\pm\,\frac{1}{~
2\para^{2}\Et(\phiX)^{2}}\,\mp\,
\frac{m_{s}^{2}}{~2\para^{2}
\left[\,H_{D0}+C_{t}(\phiX)\,\right]^{2}}\,.}
\label{nonrelLG}
\ee
Here `non-relativistic' merely refers to the fact that the kinetic term in (\ref{nonrelLG}) is
non-relativistic. The dynamics given by (\ref{nonrelLG}) is restricted to the surface of
vanishing energy in the phase space.    The equations of motion for $X$ in both systems, the DBI action
(\ref{LDBIab}) and the non-relativistic action (\ref{nonrelLG}), are equivalent,
as one can see easily by taking the time derivative of (\ref{0X}).\footnote{
However, it is  worthwhile to note that   the above
equivalence between the relativistic and  non-relativistic systems is a special property of
 one dimension. In one dimension  these  are integrable systems where
the number of dynamical variables is one, the same as the number of conserved quantities.
In higher dimensions, one might consider
\[
\ba{ll}
\displaystyle{``\cL"=
\left(\sum_{j=1}^{D}\frac{1}{2}\dot{X}^{j}{}^{2}\right)
\pm\frac{1}{~2\para^{2}\Et^{2}\,}
\mp\frac{m_{s}^{2}}{~2\para^{2}
\left(H_{D0}+C_{t}\right)^{2}}\,,}~~~&~~~
\displaystyle{X^{j}:=\frac{1}{\para}\int_{\phi^{j}_{0}}^{\phi^{j}}{\rm d}\phi^{j}{}^{\prime}\,
\frac{\Epj(\phi^{\prime})}{~
\Et(\phi^{\prime})^{2}}\,,}
\ea
\]
but in general this Lagrangian does not lead to  equations of motion
for relativistic particles.  This would require integrability.}
This corresponds  to a  canonical
transformation in  classical mechanics of the form
\be
\displaystyle{
\left(\,\phi\,,~p_{\phi}\,\right)~~\Longrightarrow~~\left(X(\phi)\,,~
P_{X}=\frac{p_{\phi}}{\,\Et\sqrt{m_{s}^{2}\Ep^{2}\pm p_{\phi}^{2}\,}}\,\right)\,,}
\ee
where  the new Hamiltonian after the transformation is given by
half of the right hand side of (\ref{0X}) with $\dot{X}\equiv P_{X}$.\\

Further we  consider the dynamics of the  D0-brane gas.  The above single particle
non-relativistic Lagrangian  has a natural generalization
to the Yang-Mills mechanics, \textit{i.e.~}to the matrix model with $\mbox{U}(N)$ gauge symmetry
for the description of $N$  D-particles,
\be
\ba{l}
\displaystyle{\cL_{{\rm M.M.}}=\cL_{\rm M.M.\,for\,DBI}~+~\cL_{\rm M.M.\,for\,CS}}\,,\\
{}\\
\displaystyle{\cL_{{\rm M.M.\,for\,DBI}}=m_{s}\,\tr\!\left[\half\left(D_{t}X\right)^{2}\,\pm\,\frac{1}{~
2\para^{2}\Et({\phiX})^{2}}\right]\,,}\\
{}\\
\displaystyle{\cL_{{\rm M.M.\,for\,CS}}=m_{s}\,\tr\!\left[\,\mp\,
\frac{1}{~2
\left[\,\cH+{\para}m_{s}^{-1} C_{t}(\phiX)\,\right]^{2}}\right]\,.}
\ea
\label{LMM}
\ee
Here $X$ is a Hermitian matrix of which the eigenvalues represent the positions of the
D-particles. Eq.(\ref{LMM}) is the generic form of the matrix model we discuss.
In the subsequent we compute the potential for different background geometries. \\

The matrix model (\ref{LMM}) describes the  D-particle dynamics. Moreover,
it is also the holographic dual of  two-dimensional 0A string theory.
As before, the dynamics is restricted to  zero energy.
The  fluctuations around  zero energy   correspond to  string excitations in
the given background.\\

In (\ref{LMM}) the ordinary  time derivative is replaced by the covariant time derivative,
with a non-dynamical  gauge field $A$,
\be
\displaystyle{D_{t}X=\dot{X}-i[A,X]\,,}
\label{DtA}
\ee
in order to allow for the gauge symmetry,
\be
\ba{lll}
X~\longrightarrow~U^{-1}XU\,,~~~~&~~~~A~\longrightarrow~U^{-1}AU+iU^{-1}\partial_{t}U\,,~~~~&~~~~
U\in\mbox{U}(N)\,.
\ea
\label{Ugauge}
\ee
The  equation of motion for the auxiliary gauge field  $A$  is
a secondary first-class constraint. This implies that
the physical states are in the gauge singlet sector.\\

There are two reasons justifying the gauging introduced in (\ref{DtA}).
The first is related to  the path integral  formulation of the
above gauged matrix model. After choosing the
diagonal gauge for $X$, the Vandermonde determinant
appears as a Faddeev-Popov determinant.
Then integrating out the gauge field $A$ precisely cancels the Vandermonde
determinant~\cite{SemenoffPrivate}. The cancellation implies
the absence of any pair interaction between  D-particles. Hence,
the D-particle is only
subject to the force originating from the background geometry as given by the
potential in (\ref{LMM}).
This is  consistent with our probe analysis approach where
 the pair interaction is negligible. \\

The second reason for the gauging is related to
the indistinguishability   of the D-particles.
The diagonalization of the $X$ is not unique; the Weyl
group of $\mbox{U}(N)$ in (\ref{Ugauge}) acts by permuting the eigenvalues of $X$.
Physically, this is just the fact that $D$-particles are identical particles,
as noted by Witten~\cite{Witten:1995im}. In  matrix models,
gauging the $\mbox{U}(N)$ symmetry  naturally
takes care of the ambiguity in the diagonalization.
Different diagonalizations correspond to the same
gauge orbit and hence to the same physical state
(see  \cite{Park:2002eu} for further discussion).
The gauging of the matrix model is also
consistent with the  gauge theory description of  $D$-brane dynamics
(see \textit{e.g.~}\cite{Suyama:2004vk}).\\

In (\ref{LMM}) we also introduce  a dimensionless parameter, $\cH$,  related to the  average  of
the D-particle energy,
\be
\displaystyle{\cH\equiv\frac{\para}{~m_{s}} \langle H_{D0}\rangle}\,.
\label{cHdef}
\ee
When the D0-branes are in  thermal equilibrium with the background geometry,
we can interpret $\cH$ as the ensemble average. Moreover, $\cH$ should be tuned
as a function of the temperature, such that
the thermodynamical   energy of the matrix model  vanishes.
This is necessary for consistency with
the fact that the dynamics is constrained to the surface of
vanishing  energy.\\

If there is no contribution from the Chern-Simons term, as in the $Q=0$ case, or
for  D-particles not charged under  the given RR flux,  \textit{e.g.}
${\rm{D}}0^{(+)}$  on $C^{(-)}$  flux,
the  matrix model $\cL_{{\rm M.M.}}$ in (\ref{LMM}) reduces to
the DBI matrix model,
\be
\displaystyle{\cL_{{\rm M.M.\,for\,DBI}}=m_{s}
\tr\!\left[\,\frac{1}{2}\left(D_{t}X\right)^{2}\,\pm\,\frac{1}{~
2\para^{2}\Et({\phiX})^{2}}\right]\,.}
\label{RRD+-}
\ee
The CS contribution  given by the last equation of (\ref{LMM}) becomes trivial,
\textit{i.e.~}depends on $\cH$ only, and the dynamics is now constrained to the
Fermi surface,
\be
\displaystyle{E_{{\rm{FS}}}\equiv\mp\frac{\,m_{s}N}{2\cH^{2}}=\mp\frac{m_{s}^{3}N}{~2 \para^{2}\langle H_{D0}\rangle^{2}}\,.}
\label{FSE}
\ee
Again the upper and lower signs refer to the Minkowskian and the Euclidean signatures.
In our conventions, $X$ has the dimension of  length, while
$\cL_{{\rm M.M.}}$ and $H_{{\rm M.M.}}$ have  mass dimension, and hence the action is dimensionless.
Since $\langle H_{D0}\rangle^{2}>0$, the Fermi surface energy  given by (\ref{FSE}) is negative
for the Minkowskian signature.\footnote{In our normalization,
$E_{{\rm{FS}}}\propto m_{s}N$. However, the overall  factor of
our matrix model is a priori arbitrary due to the restriction to  zero energy.
In  analogy  to standard  Yang-Mills theory,  we may rescale $X$ such that the
resulting Lagrangian acquires an extra factor
 $g_{s}^{-1}$. Then the Fermi sea level is proportional to
$g_{s}^{-1}$ as in \cite{McGreevy:2003kb}.}\\

\newpage

\subsection{Isometry of the background geometry, matrix model and Liouville theory
\label{ISOMM}}
The explicit form  of the DBI
potential,  $\mp\half\left[\para\Et(\phi(X))\right]^{-2}$ in (\ref{LMM}) as a function of $X$
can  in principle be obtained
after solving (\ref{Xdef}) for the inverse of $X(\phi)$.
Alternatively, using the chain rule
one can obtain  the  power series  expansion of the potential  in  $X$,
as done later in  (\ref{chainrule}).
However, remarkably,  as we  show in this subsection,
\textit{if the background geometry with the effective metric} $e^{-2\Phi}g_{\mu\nu}$
\textit{admits a nontrivial isometry, the DBI potential  must be a polynomial
 at most quadratic in} $X$.  Moreover, matrix models with at most quadratic potential,
 even with arbitrary time dependent
coefficients,  possess an $\mbox{SO}(1,2)$ symmetry as Noether symmetry~\cite{Park:2005pz}.\\

We focus on the static backgrounds with the effective metric,
\be
\displaystyle{ G_{\mu\nu}^{\Einstein}=e^{-2\Phi}g_{\mu\nu}
 =\left(\ba{cc}~\mp\Et(\phi)^{2}~&0\\0&~\Ep(\phi)^{2}~\ea\right)_{\mu\nu}\,,}
\label{NewG}
\ee
and analyze   its isometry, if any,
\be
\cL_{V}G_{\mu\nu}^{\Einstein}=0\,,
\label{ISOMG}
\ee
where $V$ is  a Killing vector. As we restrict on the static backgrounds,
there exits at least one isometry given by the time translational symmetry.
Later,  we discuss the implication of the isometry for
the corresponding matrix model.\\

For the static background, (\ref{abdef}), the Killing equations are explicitly
\be
\ba{lll}
\partial_{\phi}V^{\phi}\Ep+V^{\phi}\partial_{\phi}\Ep=0\,,~~~&~~~
V^{\phi}\partial_{\phi}\Et+\partial_{t}V^{t}\Et=0\,,~~~&~~~
\partial_{\phi}V^{t}\Et^{2}\mp\partial_{t}V^{\phi}\Ep^{2}=0\,.
\ea
\label{Killing}
\ee
It turns out that one can solve the Killing equations analytically.  The first relation implies
$V^{\phi}(t,\phi)=\Ep^{-1}(\phi)f(t)$, which further gives along other two relations,
\be
\ba{ll}
\ddot{f}(t)=\Lambda(\phi)f(t)\,,~~~~&~~~~
\displaystyle{\Lambda(\phi):=\mp\,\frac{\,\Et^{2}}{\,\Ep~}\,\partial_{\phi}\left(
\frac{\partial_{\phi}\Et}{\Et\Ep}\right)\,.}
\ea
\label{Lambdadef}
\ee
Thus, for a nontrivial solution to exist, the consistency requires
\be
\ba{lll}
\displaystyle{\frac{{\rm d}\Lambda(\phi)}{{\rm d}\phi}=0}~~~~&~~\mathrm{or}~~
&~~~~\Lambda=\mbox{constant\,.}
\ea
\ee
This is the  necessary and sufficient condition for the zweibein,  $\Et(\phi)$,
$\Ep(\phi)$, to admit any  nontrivial isometry, apart from the time translation.
Provided this condition  satisfied, one can obtain the most general solutions for  the Killing vector.
It turns out that they  are  given with three free parameters which we denote by
$c_{+}$, $c_{-}$, $c_{0}$. This implies that
the  isometry group is   three-dimensional.\\

First, when $\Lambda\neq 0$ the Killing vector  is of the general form,
\be
\ba{ll}
\displaystyle{V^{t}=\frac{\partial_{\phi}\Et}{\Et\Ep}
\Big(c_{+}f_{+}(t)+c_{-}f_{-}(t)\Big)+c_{0}\,,}~~~~&~~~~
\displaystyle{V^{\phi}=-
\frac{1}{\Ep}
\Big(c_{+}\dot{f}_{+}(t)+c_{-}\dot{f}_{-}(t)\Big)}\,,
\ea
\label{Lneq0}
\ee
where $f_{+}(t)$, $f_{-}(t)$ are the two distinct solutions of the second order differential
equation,
\be
\ddot{f}_{\pm}(t)=f_{\pm}(t)\Lambda\,.
\ee
Explicitly, for $\Lambda>0$,
\be
\ba{ll}
{f}_{+}(t)=\cosh(\sqrt{\Lambda}t)\,,~~~~&~~~~{f}_{-}(t)=\sinh(\sqrt{\Lambda}t)\,,
\ea
\ee
while for $\Lambda<0$,
\be
\ba{ll}
{f}_{+}(t)=\cos(\sqrt{\left|\Lambda\right|}t)\,,~~~~&~~~~
{f}_{-}(t)=\sin(\sqrt{\left|\Lambda\right|}t)\,.
\ea
\ee
In particular, the  parameter  $c_{0}$ amounts to
the time  translational symmetry of  the static background, corresponding to the
DBI Hamiltonian, $H_{{\rm DBI}}$.
In order to identify the   isometry algebra  we take the generators,
from (\ref{Lneq0}) for $\Lambda\neq 0$, as
\be
\ba{lll}
\displaystyle{\cJ_{+}:=-i\left(\frac{\partial_{\phi}\Et}{\Et\Ep}f_{+}\partial_{t}
-\frac{1}{\Ep}\dot{f}_{+}\partial_{\phi}\right)}\,,~&~~
\displaystyle{\cJ_{-}:=-i\left(\frac{\partial_{\phi}\Et}{\Et\Ep}f_{-}\partial_{t}
-\frac{1}{\Ep}\dot{f}_{-}\partial_{\phi}\right)}\,,~&~~
\displaystyle{H_{\rm{DBI}}:=-i\partial_{t}\,,}
\ea
\ee
and  obtain the commutator relations
\be
\ba{lll}
{}\left[\cJ_{+},\cJ_{-}\right]=-i\Omega\sqrt{\left|{\Lambda}\right|}H_{\rm{DBI}}\,,~&
{}\left[H_{\rm{DBI}},\cJ_{+}\right]=-i\,\mbox{sign}(\Lambda)\sqrt{\left|{\Lambda}\right|}
\cJ_{-}\,,~&
{}\left[H_{\rm{DBI}},\cJ_{-}\right]=-i\sqrt{\left|{\Lambda}\right|}\cJ_{+}\,.
\ea
\ee
Here $\mbox{sign}(\Lambda)$ denotes the sign of $\Lambda$, and $\Omega$ is  a constant given by
\be
\ba{ll}
\displaystyle{\Omega:=\left(\frac{\partial_{\phi}\Et}{\Et\Ep}\right)^{2}+\frac{1}{\Ep}\,
\partial_{\phi}\!\left(\frac{\partial_{\phi}\Et}{\Et\Ep}\right)\,,}~~~~&
~~~~\partial_{\phi}\Omega=0\,.
\ea
\ee
The constant property follows
from the consistency condition of the isometry (\ref{Lambdadef}). \\

Now we turn to the case  $\Lambda=0$, where the Killing vector  is of the general form
\be
\ba{ll}
\displaystyle{V^{t}=\frac{\partial_{\phi}\Et}{\Et\Ep}
\Big(\half c_{+}t^{2}+c_{-}t\Big)\mp\left(
\int\!\rmd\phi\,\frac{\cE_{\phi}}{\cE_{t}^{2}}\right)c_{+}+c_{0}\,,}~~~~&~~~~
\displaystyle{V^{\phi}=-
\frac{1}{\Ep}
\Big(c_{+}t+c_{-}\Big)}\,.
\ea
\ee
The corresponding generators are
\be
\ba{ll}
\displaystyle{\cJ^{\prime}_{+}:=-i\left[\left(
\frac{\partial_{\phi}\Et}{\,2\Et\Ep}t^{2}\mp\int\!\rmd\phi\,
\frac{\cE_{\phi}}{\,\cE_{t}^{2}\,}
\right)\partial_{t}
-\frac{t}{\,\Ep}\partial_{\phi}\right]}\,,~&
\displaystyle{\cJ^{\prime}_{-}
:=-i\left(\frac{\,\partial_{\phi}\Et}{\,\Et\Ep}t\,\partial_{t}
-\frac{1}{\Ep}\partial_{\phi}\right)}\,,
\ea
\ee
and $H_{\rm{DBI}}=-i\partial_{t}$, as before. In the case of  $\Lambda=0$,
the following two quantities are constants,
\be
\ba{ll}
\displaystyle{\frac{\partial_{\phi}\Et}{\,\Et\Ep}:=\Omega^{\prime}\,,}~~~~&~~~~
\displaystyle{
\frac{1}{\,\cE_{t}^{2}\,}+2\frac{\partial_{\phi}\Et}{\,\Et\Ep}
\int\!\rmd\phi\,\frac{\cE_{\phi}}{\cE_{t}^{2}}\,.}
\ea
\ee
Further, if the former is not zero, one can set the latter to vanish by choosing the
integration  constant properly. Accordingly, for
$\Omega^{\prime}\neq 0$  we have the commutator relations
\be
\ba{lll}
{}\displaystyle{\left[\cJ^{\prime}_{+},\cJ^{\prime}_{-}\right]
=i\Omega^{\prime}\cJ^{\prime}_{+}\,,}~&
{}\displaystyle{\left[H_{\rm{DBI}},\cJ^{\prime}_{+}\right]=-i\cJ^{\prime}_{-}\,,}~&
{}\displaystyle{\left[H_{\rm{DBI}},\cJ^{\prime}_{-}\right]=-i
\Omega^{\prime}\cJ^{\prime}_{+}\,,}
\ea
\ee
and for $\Omega^{\prime}= 0$,
\be
\ba{lll}
{}\displaystyle{\left[\cJ^{\prime}_{+},\cJ^{\prime}_{-}\right]
=\pm i\Et^{-2}H_{\rm{DBI}}\,,}~~~&~~~
{}\displaystyle{\left[H_{\rm{DBI}},\cJ^{\prime}_{+}\right]=-i\cJ^{\prime}_{-}\,,}~~~&~~~
{}\displaystyle{\left[H_{\rm{DBI}},\cJ^{\prime}_{-}\right]=0\,.}
\ea
\ee
Both of them can be identified as Poincar\'{e} algebra.\\

All together, we have the  classification of the possible isometry group
for two-dimensional static backgrounds, as summarized in {\bf{Table 1}}.
In particular, $H_{\rm{DBI}}$ corresponds to a
$\so(1,1)$ generator if  $\Lambda>0$, and $\so(2)$ generator if $\Lambda<0$.\\
\begin{table}[htb]
\begin{center}
\begin{tabular}{cccl}
\multicolumn{2}{c}{$\Lambda$, $\Omega$, $\Omega^{\prime}$} &~
&Isometry group  \\
\hline
$\Lambda>0,$ & $\Omega>0$ &: & \SO(1,2) \\
$\Lambda>0,$ & $\Omega=0$ &:& Minkowskian  Poincar\'{e} group\\
$\Lambda>0,$ & $\Omega<0$ &:& \SO(1,2)\\
$\Lambda<0,$ & $\Omega>0$ &:& \SO(1,2) \\
$\Lambda<0,$ & $\Omega=0$ &:& Euclidean   Poincar\'{e} group \\
$\Lambda<0,$ & $\Omega<0$ &:& \SO(3)\\
$\Lambda=0,$ & $\Omega^{\prime}>0$ &:& Minkowskian   Poincar\'{e} group  \\
$\Lambda=0,$ & $\Omega^{\prime}=0$ &:&
Poincar\'{e} group with the same signature as the target space\\
$\Lambda=0,$ & $\Omega^{\prime}<0$ &:& Euclidean  Poincar\'{e} group \\
\hline
\end{tabular}
\caption{Classification of possible isometry group}
\end{center}
\label{TheTable}
\end{table}

In the remaining of this subsection, we discuss the implication of the isometry to
the corresponding matrix model.
From (\ref{Xdef}), (\ref{Lambdadef}), using the chain rule, we get
\be
\ba{ll}
\displaystyle{\frac{\partial~}{\partial X}
\left(\pm\frac{1}{~2\para^{2}\Et(\phiX)^{2}}\right)=
\mp\frac{\partial_{\phi}\Et}{~\para\Et\Ep}:=\rho(\phi)\,,}~~~~&~~~~
\displaystyle{\frac{\partial^{2}~}{\partial X^{2}}
\left(\pm\frac{1}{~2\para^{2}\Et(\phiX)^{2}}\right)=\Lambda\,.}
\ea
\label{chainrule}
\ee
Remarkably this shows that \textit{when the background geometry with the effective
metric}, $e^{-2\Phi}g_{\mu\nu}$ (\ref{NewG}), \textit{admits
an isometry, apart from the time translational symmetry,  the potential in
the DBI matrix model  must be a polynomial at most quadratic in $X$,}
\be
\displaystyle{\cL_{{\rm M.M.\,for\,DBI}}\,=\tr\!\!\left[\half\left(D_{t}X\right)^{2}\pm
\Big(\!2{\para^{2}\Et^{2}}\!\Big)^{-1}\right]=
\tr\!\!\left[\half\left(D_{t}X\right)^{2}+\half\Lambda X^{2}
+\rho(\phi_{0})X+\,\mbox{constant}\,\right].}
\label{LMM2}
\ee
Moreover,  since   $\Et^{2}$ is positive definite,
$\Lambda$ must be positive  semi-definite   for  Minkowskian signature (the upper sign),
which corresponds to the upside-down potential generically.
On the other hand, for  Euclidean signature (the lower sign)
$\Lambda$ must be  negative semi-definite, and
the system describes  the usual harmonic oscillators.
 Similarly,
the ``signature'' of $H_{\rm{DBI}}$, \textit{i.e.~}whether it corresponds to $\so(1,1)$ or $\so(2)$,
is  identical to that of the background geometry.\\

The isometry naturally gives rise to the Noether symmetry of the  Dirac-Born-Infeld action.
Taking care of the gauge fixing condition $s\equiv t=x^{0}$,  one can see easily
the Noether symmetry is
\be
\delta\phi=V^{t}\dot{\phi}-V^{\phi}\,.
\ee
The corresponding Noether charge
\be
Q_{V}(\phi,p_{\phi},t)=H_{{\rm DBI}}(\phi,p_{\phi})V^{t}(t,\phi)-p_{\phi}V^{\phi}(t,\phi)\,,
\ee
generates the infinitesimal transformation via  Poisson bracket,
$\delta \phi=\left\{\phi,Q_{V}\right\}_{{\rm P.B.}}$, and satisfies
the  conservation property,
\be
\displaystyle{
\Big\{H_{\rm{DBI}}\,,\,Q_{V}\Big\}_{\rm{P.B.}}=\frac{\partial Q_{V}}{\partial t}\,.}
\ee
Due to the explicit time dependence  of the Killing vector,
$V^{t}(t,\phi)$, $V^{\phi}(t,\phi)$, the right hand side does not vanish in general.
The induced transformation of $X$ by $Q_{V}$ is
\be
{\left\{X,Q_{V}\right\}_{{\rm P.B.}}=V^{t}D_{t}{X}-V^{\phi}\Ep\Et^{-2}}\,.
\ee
However, this does not give rise to  a symmetry
for the non-relativistic Lagrangian (\ref{nonrelLG}) or
the DBI matrix model in (\ref{LMM}),  from the  following reason.
The Hamiltonian for the DBI matrix model is  related to that of the
DBI action by  $H\equiv c\half H_{{\rm{DBI}}}^{-2}$ (with $c$ some constant), so that
$Q_{V}$ is not a symmetry generator for the Hamiltonian $H$
\be
\left\{H,Q_{V}\right\}_{{\rm P.B.}}
=-c H_{{\rm DBI}}^{-3}\partial_{t}Q_{V}\neq \partial_{t}Q_{V}\,.
\ee
Nevertheless,  we can modify the charge in order to satisfy the conservation,
\be
\ba{ll}
\displaystyle{
\cQ_{V}:=Q_{V}+\sum_{n=1}^{\infty}\,\frac{~t^{n}}{n!}
\left(-1-cH_{{\rm{DBI}}}^{-3}\right)^{n}\,\frac{\partial^{n}Q_{V}}{\partial t^{n}}
~,}~~~~&~~~~\displaystyle{
\left\{H,\cQ_{V}\right\}_{{\rm P.B.}}=\partial_{t}\cQ_{V}\,.}
\ea
\ee
This guarantees  that the isometry of the background geometry is inherited by
the matrix model as a Noether symmetry.  Indeed, in \cite{Park:2005pz}
it was shown that matrix models with at most quadratic potential possess
$\so(1,2)$ symmetry, and further allow for ${\rm{osp}}(1|2,{\bf R})$
supersymmetric extension.
In Appendix  \ref{MMSO12}, we review the $\SO(1,2)$ symmetry and present
the explicit form of the transformations. \\

Finally, it is worth to note that the
symmetry algebra for the corresponding matrix model  is always
$\so(1,2)$, while the isometry of the background can be different such as
$\so(3)$ or  Poincar\'{e} algebra.
The $\so(1,2)$ symmetry  in the matrix model may be identified with  the
global subalgebra $\mbox{sl}(2,R)$ of the Virasoro algebra in string theory.
Further, the quadratic matrix models possess a
$W_{\infty}$ algebra~\cite{Park:2005pz,Winfinity}
 which   corresponds to the full Virasoro algebra.
Considering the potentially limited applicability  of the gravity action (\ref{Sbulk}) to
string theory, this might provide a criterium  for true string vacua.
In fact, if we define a new spatial  coordinate $\sigma$ and  a function
$\Phi(\sigma)$ as
\be
\ba{ll}
\displaystyle{\frac{\partial~}{\partial \sigma}:=\frac{1}{\,\Ep(\phi)}\,
\frac{\partial~}{\partial \phi}\,,}~~~~&~~~~
\displaystyle{\Phi(\sigma):=-\ln\Et(\phi)\,,}
\ea
\ee
then the defining equation of $\Lambda$ (\ref{Lambdadef}) can be rewritten as
\be
\ba{ll}
\displaystyle{\frac{\partial^{2}\Phi\,}{\partial\sigma^{2}}}=
\pm\,\Lambda \exp\!\left(2\Phi\right)\,.
\ea
\ee
Remarkably this coincides with the static equation of motion in Liouville field theory,
provided $\Lambda$ is constant which is then
identified as the Liouville background charge. This connection to the Liouville theory
supports our approach to employ the Born-Infeld  action for the
backgrounds with constant $\Lambda$.\\
~\\

\section{Simple applications\label{SimpleSec}}
In this section we apply the above formalism,  as simple exercises,   to
Rindler (\ref{Rindler}), anti-de-Sitter (\ref{AdSBH}) and
the linear dilaton (\ref{lineardilaton}) backgrounds. For the first two cases,
we freeze the dilaton at the outer horizon, $\Phi\equiv\Phi(\phi^{\outh}_{h})$,
keeping only the leading order terms in the exact  black hole solutions.
Corrections to them are considered  in the next section. All the  resulting matrix models
turn out to have   potentials at most quadratic in $X$,
and hence they have  $\SO(1,2)$ symmetry and allow for
the supersymmetric extension~\cite{Park:2005pz}.
The corresponding geometries with the effective metric
$e^{-2\Phi}g_{\mu\nu}$ admit
the isometry. \newline

\subsection{Rindler space}
Setting $\para\equiv e^{\Phi(\phi^{\outh}_{h})}$,  we get
\be
\ba{ll}
\displaystyle{\Lambda=4\pi^{2}\Tbh^{2}}\,,~~~~~&~~~~~\displaystyle{
X=\frac{1}{~4\left(\pi\Tbh\right)^{\frac{3}{2}}\sqrt{-u\,}}~,}
\ea
\ee
and hence, the DBI matrix model for the Rindler space reads
\be
\displaystyle{\cL^{{\scriptscriptstyle{\rm{DBI}}}}_{\,{\rm{Rindler}}}=m_{s}
\tr\!\left[\half\left(D_{t}X\right)^{2}\,+\,2\pi^{2} \Tbh^{2}X^{2}\right]\,.}
\label{MMRindler}
\ee
Again,   $X$ is non-negative, $X\geq 0$. The origin
$X=0$ and the infinity $X=\infty$  correspond to
the spatial infinity $u=-\infty$ and   the  horizon  $u=0$, respectively.
As follows from
\be
\displaystyle{
e^{\Phi(u)}=e^{\Phi\left(\phi_{h}^{\outh}\right)+\sqrt{2} um_{s}}
<e^{\Phi\left(\phi_{h}^{\outh}\right)}\,,}
\label{gsbound}
\ee
since $u$ is negative, the string coupling is bounded from above.
This is in contrast to the linear dilaton background,
where  the string coupling
diverges when $X\rightarrow\infty$, as we see below.\\

The matrix model corresponds to the
first  order approximation  to the near-horizon D0-dynamics for non-extremal black holes.
The corrections to the potential  coming from the higher orders as well as  the contribution from
the Chern-Simons sector are analyzed in detail  in the next section. In any case,
the leading order term in the potential
near the  horizon  or for large $X$ is the above quadratic one.
Accordingly D-branes fall with   increasing velocity  $D_{t}X\sim 2\pi\Tbh X$  and
the travel time is proportional to the logarithm of the distance
with  coefficient $\Tbh^{-1}$. For an external observer,
they are falling forever (see also \cite{Lapan:2005qz}) without  reaching the horizon:
a typical feature of black holes. \\

Since the coefficient of the quadratic  potential is proportional to $\Tbh^{2}$,
the energy spectra are given in the unit of the  temperature. Thus, the corresponding
partition function, $Z=\Tr\left(e^{-\beta_{\scriptscriptstyle{\rm BH}}E}\right)$
is independent  of the  temperature, as $\beta_{\scriptscriptstyle{\rm BH}}\Tbh=1$.
Furthermore,  the thermodynamical energy
$\langle E\rangle=\Tr\left(e^{-\beta_{\scriptscriptstyle{\rm BH}}E}E\right)Z^{-1}$
is linear  in the temperature, and
the entropy becomes independent of the temperature,
\be
\ba{ll}
\displaystyle{S=
\ln Z+\beta_{\scriptscriptstyle{\rm BH}}\langle E\rangle\,,}~~~~&~~~~
\displaystyle{\frac{\partial S~~}{\partial \Tbh}=0\,.}
\ea
\ee
However, taking into account the next order corrections to the
above matrix model for non-extremal black hole  will
make  the entropy temperature dependent. \newline
~\\
~\\

\subsection{$AdS_{2}$ space:~$AdS_{2}/\mathrm{CFT}_{1}$ correspondence}
Setting $\para\equiv e^{\Phi(\phi^{\outh}_{h})}$ and  from
\be
\ba{ll}
\displaystyle{\Lambda=0}\,,~~~~~&~~~~~
\displaystyle{m_{s}X=\frac{1}{\left(-um_{s}\right)^{2}}\,,}
\ea
\ee
the DBI matrix model for the $AdS_{2}$ space reads ({\textit{cf.}}
\cite{Strominger:2003tm,Ho:2004qp,Aharony:2005hm})
\be
\displaystyle{\cL^{{\scriptscriptstyle{{\rm{DBI}}}}}_{\,{{AdS_{2}}}}=m_{s}
\tr\!\left[\,\half\left(D_{t}X\right)^{2}\,+\,512m_{s}X\right]\,.}
\label{MMAdS2MM}
\ee
The string coupling is again bounded, as for the Rindler space (\ref{gsbound}).\\

This matrix model corresponds to the
first  order approximation  to the near-horizon dynamics of D-particles  for extremal black hole.
For large $X$,
the travel time is proportional to the square root of the distance, taking much longer than
as in the non-extremal case above.
Namely D-branes  fall much slower in the extremal black hole.
The matrix model  possesses  conformal symmetry~\cite{Park:2005pz}, (\ref{conformal1}),
\be
\ba{ll}
\delta X=\delta t \left(D_{t}X-512m_{s}t\right)\,-\,
\half\dot{\delta t}\left(X-256m_{s}t^{2}\right)\,,~~~~&~~~~
\displaystyle{\frac{{\rm{d}}^{3}\delta t}{{\rm{d}} t^{3}}=0\,.}
\ea
\label{conformalAdS2}
\ee
This corresponds to $\SO(1,2)$ symmetry and in particular, the choice $\delta t=t$
is the scale transformation. Consequently,
 the partition function is independent  of the temperature, and the thermodynamical energy is zero.
The entropy is again independent of the temperature (as in \cite{Spradlin:1999bn}).\newline
~\\

\subsection{Linear dilaton background}
For the linear dilaton background (\ref{lineardilaton}),
 with the  choice $\para\equiv g_{s}$, $\phi_{0}\equiv -\infty$,
 we obtain  $\Lambda$ and $X$  (\ref{Lambdadef}), (\ref{Xdef}),
\be
\ba{ll}
\displaystyle{\Lambda=\frac{2}{\alphap}}\,,~~~~~&~~~~~
\displaystyle{
X={{\sqrt{\frac{\alphap}{2}}}}{\,\displaystyle{e^{\sqrt{2}\phi m_{s}}\,.}}}
\ea
\label{LLDXphi}
\ee
Thus, the matrix model for the linear dilaton background reads ($\alphap=m_{s}^{-2}$)
\be
{\cL_{{\scriptscriptstyle{{\rm L.D.}}}}\,=m_{s}
\tr\!\left[~\frac{1}{2}\left(D_{t}X\right)^{2}+{{\frac{1}{\,{\alphap}}}}X^{2}
-\frac{1}{2}\cH^{-2}\,\right]\,.}
\label{LLD}
\ee
Note that the  contribution from the Chern-Simons action in (\ref{LMM})  is trivial,
\textit{i.e.~}just gives $-\half\cH^{-2}$.
Eq.(\ref{LLDXphi}) implies that   $X\geq 0$ and that
$X$ is proportional to the string coupling.
Since our analysis which uses the Born-Infeld action is valid only for small string
coupling, the matrix model (\ref{LLD}) is valid only for small $X$.
The fact that  $X\geq 0$ is similar to  the
${c}=1$ matrix model,\footnote{For  reviews of the $c=1$ matrix model see
\cite{Klebanov:1991qa,Ginsparg:1993is,Polchinski:1994mb}, and
for the D-brane interpretation see
\cite{McGreevy:2003kb,Klebanov:2003km,McGreevy:2003ep,Takayanagi:2003sm,Douglas:2003up,
McGreevy:2003dn,Gukov:2003yp,Takayanagi:2004ge,Takayanagi:2005tq,Maldacena:2005he}.} but
different from the  0A matrix
model~\cite{Douglas:2003up}, where $-\infty<X<+\infty$. \\

According to the equation of motion, the D-particles tend to move
to the strongly coupled region, $X\rightarrow +\infty$ where they  become
lighter~\cite{TakayanagiPrivate}.  In the strong coupling region
one should take into account the quantum correction to the gravity background,
and the above matrix model   may receive corrections at large $X$. This implies
that the  matrix model for the
linear dilaton background is valid for small $X$.\\

To compare  with the ${c}=1$ matrix model and the 0A matrix
model, we note that  the coefficient
of the potential differs.
In the ${c}=1$ matrix model the factor is $\frac{1}{2}\alphap^{-1}$, while
in the  0A matrix model it is $\frac{1}{4}\alphap^{-1}$.
A rescaling of  time  may resolve this discrepancy.
However,   the normalization of the time coordinate is   fixed by requiring
$\left|g_{tt}\right|\rightarrow 1$ at spatial infinity.
As we discussed before, the coefficient of the
quadratic potential is independent of the choice of $\para$ in (\ref{Xdef}).
In any case, the value of the coefficient given in (\ref{LLD}) agrees with the
conformal field theory
result~\cite{Douglas:2003up,Lapan:2005qz,Klebanov:2003km}.
There the classical trajectory $X=\cosh\!\left(\sqrt{\frac{2}{\alphap}}\,t\right)$
was obtained from the peak of the quantum wave function.
This trajectory satisfies
the equation of motion of our matrix model~(\ref{LLD}), including the precise
match of the frequency $\sqrt{\frac{2}{\alphap}}$.  \\
~\\
~\\

\section{Matrix models for black holes \label{MBHsec}}
In this section, we analyze in detail the matrix model for the non-extremal as well as the
extremal black hole geometries.
All the  matrix models obtained in the following subsections
\ref{subsectionHag}\,-\,\ref{CFT} describe D-particles
which fall towards the  horizon of a black hole. The D-particles also represent
fundamental strings. Since
the region inside the horizon is causally disconnected, the matter falling towards
the black hole - outside of it - provides an effective description of the black hole
itself.  Therefore,   we identify
the matrix models for D-particles in the black hole background
with an effective description of the black hole itself.
In principle, we could calculate  thermodynamical properties of the black hole,
\textit{e.g} the entropy, from the matrix models.
However this is beyond the scope of the present paper. \label{BEff}\\

First, we ask
if the exact black hole solution (\ref{lexact}) admits any isometry for the effective metric
$e^{-2\Phi}g_{\mu\nu}$.  We insert the solution into the expression for $\Lambda$ and get
\be
\displaystyle{\left.\alphap\Lambda(\phi)\right|_{\rm{exact}}\!=
2+\frac{~e^{2\sqrt{2}\phi m_{s}}\left(g_{s}Q\right)^{2}}{128\pi^{2}}\left[
-64\pi +e^{2\sqrt{2}\phi m_{s}}\left(32\pi \xi+
\left(g_{s}Q\right)^{2}\!\left(1-4\sqrt{2}\phi m_{s}\right)\right)\right]\,,}
\ee
which is certainly  not a constant in general. Hence,
 generic  black holes  do not admit an isometry.
In the asymptotically flat region, $\phi\rightarrow -\infty$, $\Lambda$ converges to $2$,
and  the linear dilaton background naturally possesses an isometry.\\

For the genuine black hole solution, $\Lambda$ can be reexpressed
 as a function of the distance to the outer horizon, $u=\phi-\phi_{h}^{\outh}\leq0$ and the temperature or
 $\varepsilon^{\outh}=-4\pi\sqrt{\alphap}\,{\Tbh}\leq 0$, (\ref{temp}),
\be
\ba{l}
\displaystyle{{\left.\alphap\Lambda(\phi)\right|_{\rm{exact}}}}\\
{}\\
\displaystyle{=
2-2\sqrt{2}(\varepsilon^{\outh}\!+2\sqrt{2})e^{2\sqrt{2}um_{s}}+
(\varepsilon^{\outh}\!+2\sqrt{2})
\left[\textstyle{\frac{1}{4}}(\varepsilon^{\outh}\!+6\sqrt{2})-\sqrt{2}
(\varepsilon^{\outh}\!+2\sqrt{2})um_{s}\right]\!e^{4\sqrt{2}um_{s}}}\\
{}\\
\displaystyle{=\textstyle{\frac{1}{4}}\left(\varepsilon^{\outh}\right)^{2}
+(\varepsilon^{\outh}\!+2\sqrt{2})\Big[
-4\varepsilon^{\outh}(um_{s})^{2}
-\textstyle{\frac{32}{3}}\left(\sqrt{2}\varepsilon^{\outh}+1\right)
(um_{s})^{3}+\!\cO\left[(um_{s})^{4}\right]\Big]\,.}
\ea
\ee
Note that the linear term vanishes so that
at the leading order,  there should be an isometry,
as seen in Sec.~\ref{SimpleSec}.
In particular, for the extremal case, the zeroth order is trivial,
indicating  the absence of the quadratic potential as in (\ref{MMAdS2MM}).\\

A remarkable cancellation    occurs when the black hole temperature saturates
the Hagedorn temperature, or
$\varepsilon^{\outh}=-2\sqrt{2}$, as $\Lambda$ becomes exactly  constant,
\be
\displaystyle{\Lambda=\frac{2}{\alphap}~~~~~~~~
~~~~~\mbox{if~}~~~~~\Tbh=\THag\,,}
\ee
which implies an isometry of the background and hence a quadratic potential
in the corresponding matrix model, as we will see shortly.\\

For the exact black hole solutions, we aim to calculate the matrix coordinate $X$ as a function of
the Liouville coordinate $\phi$,  or the distance from the horizon, $u=\phi-\phi_{h}^{{\outh}}$.
We focus on the  region outside of the outer horizon so that
$u\leq 0$ and $\varepsilon^{{\outh}}<0$.
We obtain
\be
\displaystyle{X}\displaystyle{=
\frac{\,g_{s}}{\,\para}\int_{\phi_{0}}^{\phi}{\rm d}\phi^{\prime}\,
\frac{\,e^{\sqrt{2}\phi^{\prime} m_{s}}}{
l(\phi^{\prime})^{\frac{3}{2}}}}\displaystyle{
=\frac{\,g_{s}e^{\sqrt{2}\phi_{h}^{{\outh}}m_{s}}}{\para m_{s}}
\int_{u_{0}m_{s}}^{um_{s}}{\rm d}y\,
\frac{e^{-2\sqrt{2}y}}{\left[
\left(\varepsilon^{{\outh}}+2\sqrt{2}\right)y-1+e^{-2\sqrt{2}y}
\right]^{\frac{3}{2}}}\,.}
\label{XBH}
\ee
Note that
$\phi\leq\phi_{h}^{\outh}$ and $y\leq 0$.
Putting $\phi_{0}\equiv -\infty$, we have $0< X<\infty$
such that $X=\infty$ corresponds to the
outer horizon $\phi=\phi_{h}^{{\outh}}$,
while $X=0$ corresponds to the asymptotically flat spatial infinity, $\phi=-\infty$.
Roughly,  the matrix coordinate $X$ stretches the near-horizon region infinitely. However,
as we do not know the exact expression for the integral,
except  the Hagedorn temperature case, \textit{i.e.~}$\varepsilon^{\outh}=-2\sqrt{2}$,
 we will  power expand  the integrand in $y$ and perform
the indefinite integral order by order. This procedure can be regarded as our regularization
scheme for $X$, as we make any integration  constant trivial.\footnote{
An alternative expansion by $e^{um_{s}}$ is available for the asymptotic region
$u\rightarrow-\infty$,
\[
\displaystyle{m_{s}X=\sum_{n=0}^{\infty}\frac{(2n+1)!}{\sqrt{2}\left(n!\right)^{2}}
\left(-4\sqrt{2}\,\Delta\right)^{-n}
e^{{\sqrt{2}(2n+1)}\Delta}\int_{-\infty}^{\sqrt{2}
\left(um_{s}-\Delta\right)}{\rm d}\eta~\eta^{n}e^{(2n+1)\eta}\,,}
\]
where we adopted (\ref{parachoice}) and
set $\Delta\equiv\left(\varepsilon^{{\outh}}+2\sqrt{2}\right)^{-1}$.
The leading term $n=0$ corresponds to (\ref{LLDXphi}) and leads to
the matrix model for the linear dilaton background.
The higher order terms then give the corrections.
However, in the present paper we focus on the near-horizon region,
and  do not pursue the analysis of this expansion.}\\

It is now natural to fix the arbitrary parameter as
\be
\displaystyle{\para\equiv g_{s}e^{\sqrt{2}\phi_{h}^{{\outh}}m_{s}}
=e^{\Phi(\phi_{h}^{{\outh}})}=4\pi{Q^{-1}}\sqrt{\sqrt{2}\left(\THag-{\Tbh}\right)\,}\,,}
\label{parachoice}
\ee
which is  the string coupling at the horizon, the region we are interested in.
Consequently we have an expression for the black hole temperature in terms of $\para Q$
\be
\displaystyle{{\Tbh}=\frac{m_{s}}{4\pi}\left|\varepsilon^{{\outh}}\right|
=\frac{m_{s}}{\sqrt{2}\,\pi}\!\left(1-
\frac{\left(g_{s}Q\right)^{2}}{{16\pi}}e^{2\sqrt{2}\phi^{{\outh}}_{h}m_{s}}\right)=
\frac{m_{s}}{\sqrt{2}\,\pi}\!\left(1-
\frac{\left(\para Q\right)^{2}}{{16\pi}}\right)>0\,.}
\label{Tbhvarep}
\ee
Further, the DBI potential is
\be
\displaystyle{-\frac{1}{~2\para^{2}\Et^{2}}=\frac{1}{~2-8\pi\left(\THag-\Tbh\right)u
-2e^{-2\sqrt{2}um_{s}}}}\,,
\ee
while  the CS potential is, with the gauge choice, $C_{t}(\phi_{h}^{\outh})\equiv 0$,
\be
\displaystyle{\frac{1}{~2\left(\cH+{\para m_{s}^{-1}C_{t}}\right)^{2}}=
\frac{1}{~2\left(\cH+
{\sqrt{8\sqrt{2}\,m_{s}^{-1}\left(\THag-{\Tbh}\right)}}\left(-um_{s}\right)\right)^{2}}\,.}
\label{ForCS}
\ee
~\\

We recall (\ref{gbound}) that unlike in the linear dilaton background,
the string coupling in the black hole background is bounded from above
by the inverse of the flux
\be
\displaystyle{e^{\Phi\left(\phi^{\outh}\right)}\leq\frac{4\sqrt{\pi}}{Q}\,.}
\ee
Thus, by taking large $Q$, the string coupling can be kept small and
quantum corrections can be neglected.
The corresponding matrix model for the black hole then
can be trusted over the entire region
up to $0< X<\infty$.\\

Using the chain rule (\ref{chainrule}),
the differentiation  of the DBI potential  reads
\be
\displaystyle{
\frac{{\rm d}~~}{{\rm d}X}\left(-\frac{1}{~2\para^{2}\Et\left(\phiX\right)^{2}}\right)
=-{2\pi}\sqrt{\left|g_{tt}\right|}e^{\sqrt{2}um_{s}}\left[
\frac{\cT(u)}{\left|g_{tt}\right|}\,+\,\THag\right]\,,}
\label{stretchedH}
\ee
where we set
\be
\displaystyle{\cT(u)\equiv{{\frac{1}{4\pi}}}\,\frac{{\rm d}
{g_{tt}(\phi)}}{{\rm d}\phi}=
\left[\Tbh-2\sqrt{2}\left(\THag-\Tbh\right)um_{s}\right]e^{2\sqrt{2}um_{s}}\,.}
\ee
Outside the outer horizon, $u$ is negative so that $\cT(u)$ is  positive.
Thus the right hand side of (\ref{stretchedH}) is  negative, and
there is no extremal point of the DBI potential outside the outer horizon, except at
the spatial infinity, $X=0$ or $u=-\infty$.  The DBI potential is monotonically decreasing
over the entire range, from zero at the spatial infinity to $-\infty$ at the outer horizon.
~\newline

\subsection{String/Black hole transition at the Hagedorn temperature:\\
Background independence in  the vanishing string coupling limit\label{subsectionHag}}
At the Hagedorn temperature, where $\varepsilon^{\outh}=-2\sqrt{2}$,
from (\ref{ForCS}) the CS matrix model becomes trivial,
\textit{i.e.~}just gives the constant term $-\half\cH^{-2}$.
The integration  for $X$ (\ref{XBH}) can be exactly performed, with $u_{0}\equiv -\infty$,
\be
\displaystyle{m_{s}X={\left({2e^{-2\sqrt{2}um_{s}}-2\,}\right)^{-\frac{1}{2}}}\,,}
\label{HBHXphi}
\ee
and hence,
\be
\displaystyle{um_{s}=-\frac{\sqrt{2}}{\,4\,}\,\ln\left[1+\frac{1}{\,2\left(m_{s}X\right)^{2}}\right]\,.}
\ee
Finally substituting this into (\ref{LMM}), we obtain the matrix model
for the black hole at the Hagedorn temperature,
\be
\ba{l}
\displaystyle{\cL^{{\rm Hagedorn}}_{{\rm Black\,Hole}}\,=m_{s}
\tr\!\left[\,\frac{1}{2}\left(D_{t}X\right)^{2}\,+\,{{\frac{1}{{\,\alphap}}}}X^{2}\,-\,
\frac{1}{2}\cH^{-2}\,\right]\,,}
\ea
\label{MMHag}
\ee
with $\cH$ as in (\ref{cHdef}) and $\alphap=m_{s}^{-2}$.
We observe that  the resulting matrix model is identical to the one
for the linear dilaton background  (\ref{LLD}).
This is a nontrivial result, since
the corresponding  geometries   (\ref{lineardilaton}) and
(\ref{HagBHsol}) are distinct. Moreover,
the expressions for the matrix variable in terms of the Liouville coordinate
(\ref{LLDXphi}) and (\ref{HBHXphi}) are different. \\

Note again that we interpret the matrix model~(\ref{MMHag}) as an
effective description of
the black hole, as discussed on page \pageref{BEff}.
Hence, we may view the fact that
(\ref{LMM}) reduces to (\ref{MMHag}) at the Hagedorn temperature
as realization of the  string/black hole transition~\cite{Susskind:1993aa,Susskind:1993ws,Susskind:2005js,
Horowitz:1996nw,Giveon:2005mi,Nakayama:2005pk,Kutasov:2005rr,Giveon:2005jv,Parnachev:2005qr}.
From $\THag-\Tbh=\frac{m_{s}}{16\sqrt{2}\pi^{2}}\left(\para Q\right)^{2}$, the transition occurs
when $\para Q$ vanishes, \textit{i.e.~}in the weak coupling limit.\newline

What  we observe here is  that when the string coupling vanishes,
there is a  degeneracy of the matrix models for different geometries,
(\ref{LLD}) and (\ref{MMHag}), \textit{i.e.} the matrix models for the
linear dilaton background and for the black hole background at the
Hagedorn temperature. Moreover when the string coupling goes to zero, the closed
string modes decouple, as
the closed string coupling  is the square of the open string coupling.
This implies that the dynamics of the D-particles, which correspond to  open strings,
becomes independent of the background geometry.
 Hence we  conjecture that
in the vanishing  string coupling limit, the D0-brane dynamics or string theory itself
 becomes  independent of the background geometry
 and the resulting matrix model is of  universal form, given by the
above matrix model (\ref{MMHag}). 
This is consistent with  universality in a thermodynamical system
at a critical point (see \textit{e.g.} \cite{Kadanoff:2000xz}).

\subsection{Near the  Hagedorn temperature\label{NearH}}
Taking $\THag-\Tbh=\frac{m_{s}}{4\pi}\left(\varepsilon^{\outh}+2\sqrt{2}\right)$
as a small parameter,
we analyze  the leading order correction to (\ref{MMHag}).
First  we expand the integrand  (\ref{XBH}) up to the linear order,
\be
\displaystyle{
\frac{e^{-2\sqrt{2}y}}{\left[
(\varepsilon^{\outh}+2\sqrt{2})y-1+e^{-2\sqrt{2}y}
\right]^{\frac{3}{2}}}\simeq\frac{e^{-2\sqrt{2}y}}{\left[-1+e^{-2\sqrt{2}y}
\right]^{\frac{3}{2}}}-\frac{3ye^{-2\sqrt{2}y}}{
~2\left[-1+e^{-2\sqrt{2}y}
\right]^{\frac{5}{2}}}\left(\varepsilon^{\outh}+2\sqrt{2}\right)
\,.}
\label{XBH2}
\ee
In this subsection, `$\,\simeq\,$' denotes  equality up to  linear order in
$\left(\varepsilon^{\outh}+2\sqrt{2}\right)$. It is convenient to introduce
\be
\displaystyle{W:=e^{-2\sqrt{2}um_{s}}-1\,.}
\ee
The matrix coordinate $X$ is, after integration from  $u_{0}\equiv -\infty$,
\be
\textstyle{m_{s}X\simeq\frac{1}{\sqrt{2}}W^{-\frac{1}{2}}+
\left[\frac{1}{4}W^{-\frac{1}{2}}-\frac{\sqrt{2}\,}{4}um_{s}W^{-\frac{3}{2}}-\frac{1}{4}\,
{\rm{arccot}}\left(W^{\frac{1}{2}}\right)\right]
\left(\varepsilon^{\outh}+2\sqrt{2}\right)\,.}
\ee
It follows that the inverse function reads
\be
\ba{ll}
\displaystyle{-um_{s}}\simeq &\displaystyle{
\frac{\sqrt{2}}{4}\ln\left[1+\frac{1}{\,2\left(m_{s}X\right)^{2}}\right]}\\
{}&{}\\
{}&\displaystyle{
~+\left[\frac{~2m_{s}X+2\left(m_{s}X\right)^{3}
\ln\left(1+\frac{1}{\,2\left(m_{s}X\right)^{2}}\right)-\sqrt{2}
\arctan\left(\sqrt{2}m_{s}X\right)}{8m_{s}X+16\left(m_{s}X\right)^{3}}
\right]\left(\varepsilon^{\outh}+2\sqrt{2}\right)\,.}
\ea
\label{umsNH}
\ee
The DBI potential is then
\be
\displaystyle{-\frac{1}{~2\para^{2}\Et^{2}}\simeq
-\left(m_{s}X\right)^{2}+\left[\frac{1}{\sqrt{2}}\left(m_{s}X\right)^{2}
-\frac{1}{2}\left(m_{s}X\right)\arctan\left(\sqrt{2}m_{s}X\right)\right]
\left(\varepsilon^{\outh}+2\sqrt{2}\right)\,,}
\ee
while  for the CS part,  due to the presence of the factor
 $\sqrt{\varepsilon^{\outh}+2\sqrt{2}}$ in (\ref{ForCS}),
 only the zeroth order in (\ref{umsNH}) contributes.\\

All together, the matrix model near the Hagedorn temperature is given by
\be
\displaystyle{\cL_{{\rm Near~Hagedorn}}=
\cL_{{\rm Near~Hag.}}^{{{\scriptscriptstyle{\rm{DBI}}}}}+
\cL_{{\rm Near~Hag.}}^{{{\scriptscriptstyle{\rm{CS}}}}}\,,}
\ee
where
\be
\ba{l}
\displaystyle{\cL_{{\rm Near~Hag.}}^{{{\scriptscriptstyle{\rm{DBI}}}}}
=m_{s}
\tr\!\left[\,\frac{1}{2}\left(D_{t}X\right)^{2}+{{\frac{1}{{\,\alphap}}}}X^{2}
+2\pi\left(\THag\!-\Tbh\!\right)X\left[\arctan\left(\sqrt{2}m_{s}X\right)-\sqrt{2}m_{s}X\right]
\,\right],}\\
{}\\
{}\displaystyle{\cL_{{\rm Near~Hag.}}^{{{\scriptscriptstyle{\rm{CS}}}}}
=m_{s}
\tr\!\left[-\frac{2}{3}\left[
{\textstyle{{\sqrt{\sqrt{2}\,m_{s}^{-1}\!\left(\THag\!-{\Tbh}\right)}}}}
\ln\left(\!{{1+\frac{1}{2\left(m_{s}X\right)^{2}}}}\!\right)
-\frac{1}{3}\cH\right]^{2}\cH^{-4}-\frac{23}{54}\cH^{-2}
\right].}
\ea
\label{MMNearHag}
\ee
We recall  that there are two types of D-particles and of RR flux. ${\rm{D}}0^{(+)}$
are charged under $C^{(+)}$ only, and ${\rm{D}}0^{(-)}$ under  $C^{(-)}$. The
above Chern-Simons part is valid for the charged D-particles.
See the discussion before (\ref{RRD+-}).\newline

\subsection{Non-extremal black hole\label{NoneB}}
Here we calculate  $X$  for the generic non-extremal black hole geometry.
The integrand in (\ref{XBH}) has the following power expansion in  $y<0$,  with
$\varepsilon^{\outh}<0$,
\be
\ba{l}
\displaystyle{
\frac{e^{\sqrt{2}y}}{\left[1+
\left((\varepsilon^{{\outh}}+2\sqrt{2})y-1\right)e^{2\sqrt{2}y}
\right]^{\frac{3}{2}}}}\\
{}\\
\displaystyle{=
\frac{1}{\,\left|\varepsilon^{\outh}\right|^{\frac{3}{2}}(-y)^{\frac{3}{2}}}\,+\,
\frac{2(\sqrt{2}\left|\varepsilon^{\outh}\right|-3)}{\left|\varepsilon^{\outh}\right|^{\frac{5}{2}}(-y)^{\frac{1}{2}}}
\,+\,\frac{\left(4\left|\varepsilon^{\outh}\right|^{2}-16\sqrt{2}
\left|\varepsilon^{\outh}\right|+30\right)}{
\left|\varepsilon^{\outh}\right|^{\frac{7}{2}}}(-y)^{\frac{1}{2}}\,+\,\cO\!\left[
\left(-y\right)^{\frac{3}{2}}\right]\,,}
\ea
\label{indefX}
\ee
and hence, after  indefinite integration, $X$ reads in terms of $u=\phi-\phi_{h}^{{\outh}}<0$,
\be
\displaystyle{m_{s}X\simeq
\frac{~2\left|\varepsilon^{\outh}\right|^{-\frac{3}{2}}}{\sqrt{-u m_{s}}}\,-\,
\frac{4(\sqrt{2}\left|\varepsilon^{\outh}\right|-3)}{\left|\varepsilon^{\outh}\right|^{\frac{5}{2}}}
\sqrt{-u m_{s}}\,-\,\frac{4\left(2\left|\varepsilon^{\outh}\right|^{2}-8\sqrt{2}
\left|\varepsilon^{\outh}\right|+15\right)}{3
\left|\varepsilon^{\outh}\right|^{\frac{7}{2}}}\left(-u m_{s}\right)^{\frac{3}{2}}}\,.
\label{Xnonex}
\ee
Here $\varepsilon^{\outh}$ as given by (\ref{Tbhvarep}) is related to the temperature of the
black hole, $y$ is the integration variable of (\ref{XBH}),
and $m_{s}$ is the string mass, $m_{s}=\alphap^{-\frac{1}{2}}$.\\

Similarly, up to  terms linear in $(-um_{s})$, we have
\be
\ba{l}
\displaystyle{\frac{1}{\,\left(m_{s}X\right)^{2}}\simeq
\frac{~\left|\varepsilon^{\outh}\right|^{3}\left(-um_{s}\right)}{4}\,,}\\
{}\\
\displaystyle{\left(m_{s}X\right)^{2}\simeq
\frac{4}{\,\left|\varepsilon^{\outh}\right|^{3}\left(-um_{s}\right)}\,-\,
\frac{\,16
\left(\sqrt{2}\left|\varepsilon^{\outh}\right|-3\right)}{
\left|\varepsilon^{\outh}\right|^{4}}\,+\,
\frac{\,32\left(2\left|\varepsilon^{\outh}\right|^{2}
-5\sqrt{2}\left|\varepsilon^{\outh}\right|+6\right)}{3\left|\varepsilon^{\outh}\right|^{5}}
\left(-um_{s}\right),}\\
{}\\
\displaystyle{\left(m_{s}X\right)^{4}\simeq
\frac{16}{~\left|\varepsilon^{\outh}\right|^{6}\left(-um_{s}\right)^{2}}\,-\,
\frac{~128
\left(\sqrt{2}\left|\varepsilon^{\outh}\right|-3\right)}{
\left|\varepsilon^{\outh}\right|^{7}\left(-um_{s}\right)}\,+\,
\frac{~256\left(8\left|\varepsilon^{\outh}\right|^{2}
-23\sqrt{2}\left|\varepsilon^{\outh}\right|+33\right)}{3\left|\varepsilon^{\outh}\right|^{8}}}\\
{}\\
\displaystyle{\qquad\qquad\qquad\,-\,
\frac{~256\left(32\sqrt{2}\left|\varepsilon^{\outh}\right|^{3}-243
\left|\varepsilon^{\outh}\right|^{2}+300\sqrt{2}\left|\varepsilon^{\outh}\right|
-240\right)}{15\left|\varepsilon^{\outh}\right|^{9}}\left(-um_{s}\right)\,.}
\ea
\label{X2ex}
\ee
For the analysis of the non-extremal black hole in this subsection,
we now take the near-horizon limit   \textit{by neglecting all the powers higher than one}:
\be
\ba{ll}
(-um_{s})^{n}\cong 0\,,~~~~&~~~~ n> 1\,.
\ea
\label{unearH}
\ee
In other words,  we focus on  length scales  up to the string
scale,\footnote{Alternatively, one may  neglect all the positive powers
restricting on the region $\left|u\right|\ll\sqrt{\alphap}$,
but it will miss all the negative powers of $X$ as well as the Chern-Simons part,
and eventually  reduce to the matrix model for the Rindler space~(\ref{MMRindler}).}
$\left|u\right|<\frac{1}{\Ncut}\sqrt{\alphap}$, where the cutoff is controlled  by
a natural number $\Ncut$ of order  one.\footnote{Note that unlike in higher dimensions,
in the two-dimensional near-horizon limit, $\alphap$ may remain finite.}
Therefore the expansion is valid for $X$ in the range
\be
\displaystyle{X~\gtrsim~ X_{\rm min}\equiv\frac{\sqrt{m_{s}\Ncut}}{\,4\left(\pi\Tbh\right)^{\frac{3}{2}}}
\,.}
\label{Xmin}
\ee

From (\ref{X2ex}), we can now solve for $\left(-um_{s}\right)$, $\left(-um_{s}\right)^{-1}$ and
$\left(-um_{s}\right)^{-2}$ in terms of $X$,
\be
\ba{l}
\displaystyle{\left(-um_{s}\right)\cong\frac{4}{~\left|\varepsilon^{\outh}\right|^{3}
\left(m_{s}X\right)^{2}}\,,}\\
{}\\
\displaystyle{\frac{1}{\left(-um_{s}\right)}\cong
\frac{~\left|\varepsilon^{\outh}\right|^{3}}{4}\left(m_{s}X\right)^{2}
\,+\,\frac{~4
\left(\sqrt{2}\left|\varepsilon^{\outh}\right|-3\right)}{
\left|\varepsilon^{\outh}\right|}\,-\,
\frac{~32\left(2\left|\varepsilon^{\outh}\right|^{2}
-5\sqrt{2}\left|\varepsilon^{\outh}\right|+6\right)}{3\left|\varepsilon^{\outh}\right|^{5}
\left(m_{s}X\right)^{2}}\,,}\\
{}\\
\displaystyle{\frac{1}{\left(-um_{s}\right)^{2}}\cong
\frac{~\left|\varepsilon^{\outh}\right|^{6}}{16}\left(m_{s}X\right)^{4}
\,+\,2\left(\sqrt{2}\left|\varepsilon^{\outh}\right|-3\right)
\left[\left|\varepsilon^{\outh}\right|^{2}\left(m_{s}X\right)^{2}\,+\,
\frac{~8
\left(2\sqrt{2}\left|\varepsilon^{\outh}\right|-7\right)}{
3\left|\varepsilon^{\outh}\right|^{2}}\right]}\\
{}\\
\displaystyle{\qquad\qquad\qquad\,-\,
\frac{~64\left(\left|\varepsilon^{\outh}\right|-2\sqrt{2}\right)
\left(8\sqrt{2}
\left|\varepsilon^{\outh}\right|^{2}-45\left|\varepsilon^{\outh}\right|
+30\sqrt{2}\right)}{15\left|\varepsilon^{\outh}\right|^{6}
\left(m_{s}X\right)^{2}}\,.}
\ea
\label{uexpression}
\ee
Accordingly   we have for the DBI potential,
\be
\ba{ll}
\displaystyle{\frac{1}{~\para^{2}\Et^{2}}}&
\displaystyle{=e^{2\sqrt{2}um_{s}}\left[1-\left(1+(\left|\varepsilon^{\outh}\right|-2\sqrt{2}
)um_{s}\right)
e^{2\sqrt{2}um_{s}}\right]^{-1}}\\
{}&{}\\
{}&\displaystyle{\cong \frac{1}{~\left|\varepsilon^{\outh}\right|\left(-um_{s}\right)}\,-\,
\frac{4}{~\left|\varepsilon^{\outh}\right|^{2}}\,-\,\frac{~8\left(\frac{\sqrt{2}}{3}
\left|\varepsilon^{\outh}\right|-2\right)}{\left|\varepsilon^{\outh}\right|^{3}}
\left(-um_{s}\right)}\\
{}&{}\\
{}&\displaystyle{\cong\frac{\,\left|\varepsilon^{\outh}\right|^{2}}{4}
\left(m_{s}X\right)^{2}\,+\,\frac{~4\sqrt{2}
\left(\left|\varepsilon^{\outh}\right|-2\sqrt{2}\right)}{\left|\varepsilon^{\outh}\right|^{2}}\,
-\,\frac{~64\left(\left|\varepsilon^{\outh}\right|-2\sqrt{2}\right)}{3
\left|\varepsilon^{\outh}\right|^{5}\left(m_{s}X\right)^{2}}\,.}
\ea
\ee
From (\ref{Hagedorn}) and (\ref{Tbhvarep}),
$\left|\varepsilon^{{\outh}}\right|$ corresponds to the black hole temperature and
$2\sqrt{2}$ to  the Hagedorn temperature.\\

The analysis of the near-horizon limit of the  CS matrix model is a bit more subtle due to the
arbitrary parameter $\cH$.  From (\ref{ForCS}), there is a
critical value of $\cH$, depending on the temperature and the cutoff,
\be
\displaystyle{
\cH_{\rc}(\Tbh)\equiv{\frac{1}{\Ncut}\sqrt{8\sqrt{2}\,m_{s}^{-1}\left(\THag-{\Tbh}\right)}}\,.}
\label{Hmin}
\ee
When $\left|\cH\right|>\cH_{\rc}$, the CS matrix model has a regular expansion in  $um_{s}$
and in the near-horizon region (\ref{unearH}), we need to keep only the linear term, leading to
a $X^{-2}$ potential from (\ref{uexpression}). On the other hand, when
$0<\left|\cH\right|\leq\cH_{\rc}$, the expansion depends  on
whether $\left|um_{s}\right|$ is bigger or smaller than $\left|\cH\right|$.
Therefore there is no closed expression for the  matrix model,
valid in the entire near-horizon region.
Hence we exclude the parameter range $0<\left|\cH\right|\leq\cH_{\rc}$,
as far as the Chern-Simons part is concerned. Finally when
$\cH=0$, from (\ref{ForCS}) the  CS potential is of the form $\left(um_{s}\right)^{-2}$.
From (\ref{uexpression}) this gives rise to a potential involving
$X^{4}$, $X^{2}$ and $X^{-2}$ terms.\\


All together,   the matrix model for the D-particles in the near-horizon region of the
non-extremal black hole reads
\be
\displaystyle{\cL_{\,{\rm{Black\,Hole}}}=\cL^{{{\scriptscriptstyle{\rm{DBI}}}}}_{\,{\rm{Black\,Hole}}}\,+\,
\cL^{\scriptscriptstyle{{\rm{CS}}}}_{\,{\rm{Black\,Hole}}}\,,}
\label{MMnonextBH}
\ee
where
\begin{eqnarray}
{}&&\displaystyle{\cL^{{{\scriptscriptstyle{\rm{DBI}}}}}_{\,{\rm{Black\,Hole}}}=m_{s}
\tr\!\left[\half\left(D_{t}X\right)^{2}\,+\,2\pi^{2} \Tbh^{2}X^{2}
\,+\left(\!
\frac{{\left(\THag\!-{\Tbh}\right)}m_{s}^{2}}{24\pi^{4}\Tbh^{5}}\right)\!
\frac{1}{\,X^{2}}\,-\,
\frac{{\left(\THag\!-{\Tbh}\right)}m_{s}}{\sqrt{2}\,\pi \Tbh^{2}}\right]}\,,\nonumber\\
{}&&\label{DBInonextr}\\
{}&&\displaystyle{\cL^{{{\scriptscriptstyle{\rm{CS}}}}}_{\,{\rm{Black\,Hole}}}=m_{s}
\tr\!\left[\frac{\sqrt{2\sqrt{2}m_{s}\left(\THag-{\Tbh}\right)}}{8\pi^{3}\cH^{3}\Tbh^{3}X^{2}}
~-\frac{1}{\,2\cH^{2}}\right]\,.}
\label{cHneq0}
\end{eqnarray}
Compared to the matrix model for  Rindler space (\ref{MMRindler}), the above DBI matrix model
contains an additional  $X^{-2}$ contribution to the potential.
In fact, the inverse square potential was proposed
by Jevicki and  Yoneya  for the matrix description of a two-dimensional black hole, the
so-called  deformed matrix model~\cite{Jevicki:1993zg}.\\

A few other comments are in order: The  matrix model is valid over
the range $X_{\rm min}<X$ (\ref{Xmin}),   and
$\cH_{\rc}<\left|\cH\right|$ (\ref{Hmin}) for the CS sector.
If $\cH$ vanishes,   the CS potential can be read off from (\ref{ForCS}).
The ${X^{-2}}$ repulsive potential in the DBI matrix model has an
analogy with  the known relativistic ``centrifugal" barrier, which
is dominant far from the horizon but suppressed by the gravitational
attraction $X^{2}$ near the horizon (see \textit{e.g.}
\cite{Susskind:2005js}). On the other hand, the inverse square
potential in the CS matrix model can be either repulsive or
attractive depending on the sign of $\cH$, $\cH_{\rc}<\cH$ or
$\cH<-\cH_{\rc}$ respectively.  At the
Hagedorn temperature the  above matrix model again reduces to the matrix
model for the linear dilaton  background (\ref{LLD}).
Hence our small $um_{s}$ calculation of this subsection is consistent with the exact result of
 Sec.~\ref{subsectionHag}. \\

Combining the DBI and CS parts, the total potential is
\be
\displaystyle{
V_{\rm total}=m_{s}\tr\!\left[-2\pi^{2} \Tbh^{2}X^{2}
\,-\left(\!
\frac{{\left(\THag\!-{\Tbh}\right)}m_{s}^{2}}{24\pi^{4}\Tbh^{5}}+
\frac{\sqrt{2\sqrt{2}m_{s}\left(\THag-{\Tbh}\right)}}{8\pi^{3}\cH^{3}\Tbh^{3}}\right)\!
\frac{1}{\,X^{2}}
\right]\,,}
\ee
so that depending on the value of $\cH$, the potential develops an infinite wall or  cliff
near the origin. The former will  surely prevent  the D-particles from
approaching the forbidden  region, $0\leq X<X_{\rm min}$, and the matrix model can be
trusted over the entire region $0\leq X\leq \infty$. Explicitly,  this is the case
when $\cH$  is negative and satisfies
\be
\displaystyle{
\cH_{\rc}
\,<\,-\cH\,<\,
\left(\frac{18\sqrt{2}\pi^{2}\Tbh^{4}}{m_{s}^{3}\left(\THag-\Tbh\right)}\right)^{\frac{1}{6}}\,.}
\ee
From (\ref{Hmin}) the consistency requires that
\be
\displaystyle{\frac{\THag}{
~1+\sqrt{3\pi}\left({\textstyle{\frac{1}{8}}}
\Ncut\right)^{\frac{3}{2}}}~<~\Tbh~\leq~\THag\,.}
\ee
In particular, the above condition can  always be  satisfied near  the Hagedorn temperature,
where $\para Q$ is small. In this case, the matrix model becomes
\be
\displaystyle{\cL_{\,{\rm{Black\,Hole}}}\simeq m_{s}
\tr\!\left[\half\left(D_{t}X\right)^{2}\,+\,\frac{1}{\alphap}X^{2}\,+\,
\,\alphap\left(1+\frac{12}{\,\cH^{3}\para Q\,}\right)\!
\frac{\left(\para Q\right)^{2}}{~96\pi X^{2}}\right]}\,,
\ee
which agrees with (\ref{MMNearHag}),   up to a  shift of
$X$ since  we performed  an indefinite integration for (\ref{indefX}).
The Fermi sea level is  $E_{{\rm FS}}=-\frac{1}{16\pi}
\left(\para Q\right)^{2}-\half\cH^{-2}$.
With a suitable negative choice of $\cH$, such that $\cH^{-3}\propto-\para Q$,
this Lagrangian can be identified
with the 0A matrix model proposed in \cite{Douglas:2003up,Maldacena:2005he}.\\
~\\
~\\


\subsection{Extremal black hole\label{subsectionextr}}
For the extremal black hole we have $\varepsilon=0$, and
the integrand in (\ref{XBH}) has a  regular pole. Consequently  $X$ has an
expansion different from (\ref{Xnonex}),
\be
\displaystyle{X=
\frac{\sqrt{\pi}}{~4\para m_{s}Q~}\left[\frac{1}{\left(-u m_{s}\right)^{2}}
+\frac{2\sqrt{2}}{\left(-u m_{s}\right)}-\frac{\,4\ln\left(-u m_{s}\right)}{3}
+\frac{8\sqrt{2}\left(-um_{s}\right)}{135}
\,+\,\cO\!\left[
\left(-um_{s}\right)^{2}\right]\right]}\,.
\label{extrX}
\ee
We take the near-horizon limit   by neglecting all the terms which vanish when
$um_{s}\rightarrow 0$ or
\be
\ba{lll}
(-um_{s})^{n}\cong 0\,,~~~~&~~~~ (-um_{s})^{n}\ln\left(-um_{s}\right)\cong 0\,,
~~~~&~~~~ n> 0\,.
\ea
\label{nearHextr}
\ee
Accordingly, after setting
\be
\displaystyle{\para\equiv e^{\Phi(\phi^{\outh}_{h})}=
4\sqrt{\pi}Q^{-1}\propto g_{s}\,,}
\ee
we get
\be
\ba{ll}
\displaystyle{4\sqrt{m_{s}X}\cong\frac{1}{\left(-um_{s}\right)}+\sqrt{2}\,,}~~~~&~~~~
\displaystyle{\ln\left(16m_{s}X\right)\cong -2\ln\left(um_{s}\right)\,,}
\ea
\ee
so that with (\ref{extrX})
\be
\displaystyle{{\left(-um_{s}\right)^{-2}}\cong 16m_{s}X-8\sqrt{2m_{s}X}-
{\textstyle{\frac{2}{3}}}\ln\left(16m_{s}X\right)+4\,.}
\label{uexpressionext}
\ee
The DBI potential becomes
\be
\ba{ll}
{\displaystyle{-\frac{1}{\,2\para^{2}\Et(\phiX)^{2}}}}&
{\displaystyle{\cong -{32}\left(-um_{s}\right)^{-2}
-{\textstyle{\frac{64\sqrt{2}}{3}}}\left(-um_{s}\right)^{-1}-{\textstyle{\frac{64}{9}}}}}\\
{}&{}\\
{}&{\displaystyle{\cong -512m_{s}X+{\textstyle{\frac{512}{3}}}\sqrt{2m_{s}X\,}
+{\textstyle{\frac{64}{3}}}\ln\left(16m_{s}X\right)-{\textstyle{\frac{832}{9}}}\,.}}
\ea
\ee
On the other hand for the CS potential we have $
\left(\cH+\para C_{t}m_{s}^{-1}\right)^{-2}=\left(\cH+
\sqrt{\frac{8}{\pi}}\left|um_{s}\right|\right)^{-2}$, which has a regular expansion for
 $\left|\cH\right|>\sqrt{\frac{8}{\pi}}\left|um_{s}\right|$,
and hence becomes  trivial in the near-horizon limit (\ref{nearHextr}).
On the other hand, the near-horizon limit justifies us to focus on the parameter range
 $\left|\cH\right|>\sqrt{\frac{8}{\pi}}\left|um_{s}\right|$ or $\cH=0$,
as done for the non-extremal case. When  $\cH=0$, we have $(-um_{s})^{-2}$ for the CS potential
of which   the expression   can be read off from (\ref{uexpressionext}).
For large $\left|\cH\right|$  we conclude that the matrix model for  D-particles in the near-horizon region  of
the extremal black hole is  of the form
\be
\displaystyle{\cL_{\,{\rm{extr.\,B.H.}}}
=\tr\!\left[\,\half\left(D_{t}X\right)^{2}+
512m_{s}X-{\textstyle{\frac{512}{3}}}\sqrt{2m_{s}X\,}
-{\textstyle{\frac{64}{3}}}\ln\left(16m_{s}X\right)+{\textstyle{\frac{832}{9}}}-\half\cH^{-2}\,
\right]\,.}
\label{MMextBH}
\ee
The leading order linear contribution to the potential corresponds to the exact $AdS_{2}$
background (\ref{MMAdS2MM}), while the other give corrections necessary
for the extremal black hole.
The correction terms break the conformal symmetry and the partition function
is no longer temperature independent. It will be interesting to calculate the entropy of the
above matrix model and to compare with the gravity result.\newpage

\subsection{Conformal field theory background\label{CFT}}
Here we consider the   black hole background
in a coset conformal field theory at level $\kappa$ (\ref{kmetric})
\be
\ba{ll}
\displaystyle{\rmd
s^{2}=\kappa\alphap\left(-\tanh^{2}\!\rho\,\rmd \tau^{2}+\rmd\rho^{2}\right)\,,}
~~~~&~~~~\displaystyle{\Phi=-\ln\cosh\rho +\Phi_{h}\,.}
\ea
\label{kmetric1}
\ee
When $\kappa={1/2}$,  this geometry  coincides with
the black hole  at the Hagedorn temperature
(\ref{HagBHsol}), as shown in (\ref{kmetric2}),
while for generic  $\kappa\neq {1/2}$, it is not a classical
solution of the gravity action (\ref{Sbulk}).
The corresponding matrix model is then
\be
\ba{l}
\displaystyle{\cL_{\kappa\,{\rm{Black\,Hole}}}\,=m_{s}
\tr\!\left[\,\frac{1}{2}\left(D_{t}X\right)^{2}\,
+\,{{\frac{1}{{\,2\kappa\alphap}}}}X^{2}\,-\,
\frac{1}{2}\cH^{-2}\,\right]\,.}
\ea
\label{MMkappa}
\ee
This  coincides with the matrix model for the Rindler space (\ref{MMRindler}) if we
identify the black hole temperature with
\be
\Tbh=\frac{1}{\,2\pi\sqrt{\kappa\alphap}}\,.
\label{Tkappa}
\ee
This  agrees with the conformal field theory analysis~\cite{Nakayama:2005pk}.
Especially, when $\kappa={1/2}$, we have precisely the same
matrix model as for the black hole at the Hagedorn temperature~(\ref{MMHag}).
As was discussed in Sec.\ref{subsectionHag}, this also coincides with the matrix model for
the linear dilaton background. Our interpretation of this result as
the black hole/string transition  agrees with the conformal
field theory analysis \cite{Nakayama:2005pk},  where
  $\kappa=1/2$ was shown to correspond to the far extreme of the string
 phase.
~\\
\newline
\begin{center}
\large{\textbf{Acknowledgments}}
\end{center}
We thank Johannes Gro{\ss}e and Dimitrios Tsimpis for helpful discussions.
JHP wishes to thank Tadashi Takayanagi for useful electronic  correspondence.
JHP and CS are grateful to the Alexander von Humboldt foundation for financial support.

\newpage
\appendix


\section{Matrix models with  $\mbox{SO}(1,2)$ and gauge symmetries\label{MMSO12}}
Most of the matrix models we consider have $\SO(1,2)$ symmetry.
Here we discuss how the symmetry is realized.\\

To begin, we review the $\so(1,2)$ structure behind a second order differential
equation~\cite{Park:2005pz}.  For a  given arbitrary
time dependent function, $\Lambda(t)$, we consider the following second order
differential equation,
\be
\displaystyle{\ddot{f}(t)=f(t)\Lambda(t)}\,.
\ee
Naturally there are two distinct solutions and  we denote them by  $f_{+}(t)$ and $f_{-}(t)$.
Since
\be
\displaystyle{
\frac{{\rm d}~}{{\rm d}t}\left(f_{+}\dot{f}_{-}-f_{-}\dot{f}_{+}\right)=0\,,}
\ee
if we set a non-zero constant,
\be
\displaystyle{c:=f_{+}(t)\dot{f}_{-}(t)-f_{-}(t)\dot{f}_{+}(t)\neq 0},
\ee
and define
\be
\ba{lll}
J_{0}:=-i\frac{1}{\,2c\,}
\left(f_{+}^{2}+f_{-}^{2}\right)\partial_{t}\,,~~~&~~~
J_{1}:=-i\frac{1}{\,2c\,}\left(f_{+}^{2}-f_{-}^{2}\right)\partial_{t}\,,~~~&~~~
J_{2}:=-i\frac{1}{\,c\,}f_{+}f_{-}\,\partial_{t}\,,
\ea
\label{Jso(1,2)}
\ee
then the
$\mathbf{sp}(2,{\mathbf{R}})\equiv\so(1,2)\equiv\mathbf{sl}(2,{\mathbf{R}})$ Lie algebra follows
in the standard form
\be
\ba{lll}
{}\left[J_{0},J_{1}\right]=+iJ_{2}\,,~~~~&~~~~{}\left[J_{1},J_{2}\right]=-iJ_{0}\,,~~~~&~~~~
{}\left[J_{2},J_{0}\right]=+iJ_{1}\,.
\ea
\ee
It is useful to note that the induced infinitesimal transformations, $\delta t$, generated by
$iJ_{k}$  can be equivalently specified as the three
general solutions of the following third order differential equation,
\be
\displaystyle{\frac{{\rm d}^{3}\delta t}{{\rm d}t^{3}}=4
\Lambda\frac{{\rm d}\delta t}{{\rm d}t}+2\frac{{\rm d}\Lambda}{{\rm d}t}\delta t\,.}
\label{3rdD}
\ee
~\\

Further, we introduce one more arbitrary time dependent function $\rho(t)$, and consider the
following second order differential equation,
\be
\displaystyle{\ddot{\chi}(t)=\Lambda(t)\chi(t)+\rho(t)\,.}
\label{chidef}
\ee
In order to write down the general solution explicitly in terms of $f_{\pm}(t)$, it is
convenient to define, with arbitrary  constants, $\kappa_{\pm}(0)$,
\be
\ba{ll}
\displaystyle{
\kappa_{+}(t):=\kappa_{+}(0)+\int_{0}^{t}{\rm d}t^{\prime}\rho(t^{\prime})f_{+}(t^{\prime})\,,}
~~~~&~~~~
\displaystyle{
\kappa_{-}(t):=\kappa_{-}(0)+\int_{0}^{t}{\rm d}t^{\prime}\rho(t^{\prime})f_{-}(t^{\prime})\,.}
\ea
\ee
Then the most general solution reads explicitly,
\be
\displaystyle{\chi(t)=\frac{1}{c}
\Big(f_{-}(t)\kappa_{+}(t)-f_{+}(t)\kappa_{-}(t)\Big)\,.}
\label{chidefp}
\ee
The arbitrary constants, $\kappa_{\pm}(0)$,
amount to the homogeneous solutions to (\ref{chidef}).\\

Now we consider the following  $N\times N$ matrix model with $\mbox{U}(N)$ gauge symmetry,
\be
\displaystyle{
{\cal L}_{\rm{SO}(1,2)}=\tr\!\left[\half\left(D_{t}X\right)^{2}
+\half {\Lambda}(t)X^{2}+\rho(t)X+\frac{2\const}{\left(X-\chi(t){{ 1}}\right)^{2}}\right]\,,}
\label{BosonicM}
\ee
where $\const$ is a  constant, and
$\Lambda(t)$,  $\rho(t)$ are  arbitrary time dependent coefficients,
while  $\chi(t)$ is \textit{a}
solution of the second order differential equation~(\ref{chidef}). \\

The crucial fact is that there exits  $\SO(1,2)$  symmetry
in the matrix model generated  by the following  infinitesimal transformation,\footnote{In general,
$\delta A$ can be an arbitrary linear combination of $D_{t}X$ and $X$, and here for simplicity we put
 $\delta A\equiv 0$. However, in the supersymmetric
extension of the above matrix model it is required to set $\delta X=\delta A$ and $\const=0$
\cite{Park:2005pz}.}
\be
\ba{ll}
\delta X=\delta t D_{t}X\,-\,\half\dot{\delta t}X\,-\,\zeta(t) 1\,,
~~~~~&~~~~~\delta A=0\,.
\ea
\label{conformal1}
\ee
The  diffeomorphism,  $\delta t$, is generated by $iJ_{k}$
(\ref{Jso(1,2)}) above,
or equivalently as general solutions of the third order differential equation (\ref{3rdD}).
The inhomogeneous  term is  given for each $\delta t$ by
\be
\displaystyle{\zeta(t)=\delta t\dot{\chi}(t)
-\half\dot{\delta t}\chi(t)\,.}
\ee
$X$ is a quasi-primary operator of the ``conformal" weight ${1/2}$.\\

Especially, in the absence of the inverse square potential, \textit{i.e.~}$\const=0$, we do not need
to fix $\chi(t)$ as a solution of the second order differential equation (\ref{chidef}).
Consequently there are two parameter freedom in the inhomogeneous  term,
which  amounts to the following
extra symmetries,\footnote{Hence, when $\const=0$, the  inhomogeneous  term can be alternatively
given as the general solutions of
\[
\ddot{\zeta}=\Lambda \zeta+\textstyle{\frac{3}{2}}\rho\dot{\delta t}
+\dot{\rho}\,\delta t\,,
\]
and the extra symmetries (\ref{extra}) correspond to the homogeneous part of the
solutions~\cite{Park:2005pz}.}
\be
\ba{lll}
\delta X=f_{+}(t)1~~~~&~~~~\mbox{and}~~~~&~~~~\delta X=f_{-}(t)1\,.
\ea
\label{extra}
\ee
Turning on the inverse square potential, $\const\neq 0$,  these extra symmetries are all broken.\\

It is worth to note that when $\Lambda$ and $\rho$ are constants,  we have
$\chi(t)=-{\rho}{\Lambda}^{-1}+af_{+}(t)+bf_{-}(t)$ with two arbitrary constants
$a$, $b$. Only if these  vanish, the system corresponds to the constant shift of $X$ by
${\rho}{\Lambda}^{-1}$ from the $\rho=\chi=0$ system.
If $\Lambda$ is a positive constant and $\rho=\chi=0$, it corresponds to
the `deformed matrix model'~\cite{Jevicki:1993zg};
if $\Lambda$ is a positive constant and $\rho=\const=0$, it amounts to
the $\so(1,2)$ subalgebra of the $W_{\infty}$  algebra~\cite{Winfinity};
if $\Lambda=0$, then $\delta t$ is quadratic in $t$, and  the $\SO(1,2)$ symmetry can be identified as a conformal symmetry.
With one more condition $\rho=0$, it reduces  to the well known
 conformal matrix model~\cite{deAlfaro:1976je}.\\

Introducing an auxiliary matrix, $F$, we can rewrite the  Lagrangian~(\ref{Lagrangian0})
in a non-singular manner,
\be
{\cal L}^{\prime}_{{\rm SO}(1,2)}=\tr\Big[\half\left(D_{t}X\right)^{2} +\half
\left({\Lambda}(t)-F^{2}\right)X^{2}+\left(\rho(t)+\chi(t)F^{2}\right)X-\half\chi(t)^{2}F^{2}
+2\sqrt{\const}F\,\Big]\,.
\label{Lagrangian0}
\ee
~\\

However, our analysis in Sec.~\ref{ISOMM}
shows that the inverse square potential is absent in the matrix model
formulation of the DBI action  in any isometric background.


\end{document}